\documentclass[a4paper,12pt]{article}
\usepackage{amsmath}
\usepackage{accents}
\usepackage{bm}
\usepackage[english]{babel}
\usepackage{graphicx}
\usepackage[usenames]{color}
\usepackage{colortbl}
\usepackage[font=footnotesize,figurewithin=none]{caption}
\usepackage{subcaption}
\usepackage{floatrow}
\begin{document}

\centerline {\LARGE{Detecting the Lee-Yang zeros of a high-spin system}}
\centerline {\LARGE{by the evolution of probe spin}}
\medskip
\centerline {A. R. Kuzmak$^1$, V. M. Tkachuk$^2$}
\centerline {\small \it E-Mail: andrijkuzmak@gmail.com$^1$, voltkachuk@gmail.com$^2$}
\medskip
\centerline {\small \it Department for Theoretical Physics, Ivan Franko National University of Lviv,}
\medskip
\centerline {\small \it 12 Drahomanov St., Lviv, UA-79005, Ukraine}

{\abstract
Recently in paper [Peng et al., Phys. Rev. Lett. {\bf 114}, 010601 (2015)] the experimental observation of the Lee-Yang zeros of
an Ising-type spin-$1/2$ bath, by measuring the coherence of a probe spin, was reported. We generalize this problem to the case of
an arbitrary high-spin bath. Namely, we consider the evolution of a probe arbitrary spin which interacts with bath composed
by $N$ arbitrary spins. As a result, the connection between the observed values of the probe spin, such as magnetization and
susceptibility, and the Lee-Yang zeros is found. We apply these results to some models, namely, a triangle spin cluster, the Ising model with a long-range
interaction and the 1D Ising model with nearest-neighbor interaction. Also we propose the implementation of these models on real physical systems.

\medskip

}

\section{Introduction}

Lee, Yang and Fisher in their works \cite{LeeYangZeros1,LeeYangZeros,FisherZeros} discovered a new way of studying the thermodynamics properties
of different systems \cite{PYSM} which is based on an analysis of the partition function zeros. Zeros of the partition function define the
analytic properties of free energy and thus are used as a tool for studying the nature of phase transitions. Lee and Yang considered the general
ferromagnetic Ising model with Hamiltonian
\begin{eqnarray}
H_{b}=-\sum_{i,j}J_{ij}s_i^{z}s_j^{z}-h\sum_{i}s_i^{z},
\label{bathham}
\end{eqnarray}
where $s_i^{z}$ is the $z$ component of the spin-$1/2$ operator, $J_{ij}\geq 0$ are the interaction couplings and $h$ is the value of the magnetic field.
Note that instead of the Ising model the isotropic Heisenberg model with interaction $\left(-\sum_{i,j}J_{ij}{\bf s}_i{\bf s}_j\right)$ or some other interactions
that commute with total spin of system can be used. The partition function of system (\ref{bathham}) at temperature $T$ can be expressed as an $N$th
polynomial of $z\equiv\exp{\left(-\beta h\right)}$ as
\begin{eqnarray}
Z\left(\beta, h\right)=\textrm{Tr}\left[e^{-\beta H_{b}}\right]=e^{\beta Nsh}\sum_{n=0}^{2Ns}p_{n}z^{n},
\label{partfunc}
\end{eqnarray}
where $p_n$ is the partition function with zero magnetic field under the constraint that the total
spin of the system has a projection $Ns-n$, $\beta=1/T$ is the inverse temperature and $s$ is the value of each spin in the bath which equals $1/2$.
Here, the Boltzmann constant is put unity. For specific $z$ this partition function becomes zero. These $z$ are called Lee-Yang zeros \cite{LeeYangZeros} and we denote them as $z_n$ with
$n=1,2,\ldots,2Ns$. They are located in the complex plane which corresponds to complex parameters of the Hamiltonian.
So, in \cite{LeeYangZeros} Lee and Yang proved the theorem that the zeros of partition function of the ferromagnetic Ising model are lying on the unit circle
in the complex plane $z$. Therefore, we can write $z_n=e^{i\theta_n}$. Then, partition function (\ref{partfunc}) can be expressed as follows
\begin{eqnarray}
Z\left(\beta, h\right)=p_0e^{\beta Nsh}\prod_{n=1}^{2Ns}\left(z-z_n\right).
\label{partfunc2}
\end{eqnarray}
In later works this theorem was generalized to the ferromagnetic Ising model of an arbitrary high spin \cite{genforarbspin1,genforarbspin2,genforarbspin3}
and other types of interaction \cite{genforarbint1,genforarbint2,genforarbint3,genforarbint4} including the ferromagnetic anisotropic Heisenberg model
\cite{LeeYangZerosahi}. However, for others many-body systems, the Lee-Yang zeros are not always distributed along the unit circle. Also, it is worth noting
that the Lee-Yang zeros can be generalized as zeros of partition function with respect to other physical parameters. For instance, Fisher considered
zeros of partition function with a complex temperature \cite{FisherZeros}.

The zeros of partition function of different many-body systems such as spin systems (see, for example,
\cite{zerospartfuncspin1,zerospartfuncspin2,zerospartfuncspin3,zerospartfuncspin4,zerospartfuncspin5,zerospartfuncspin6,zerospartfuncspin7} and references therein),
Bose (see, for example, \cite{zerospartfunbose1,zerospartfunbose2,zerospartfunbose3,zerospartfunbose4,zerospartfunbose5} and Fermi
(see, for example, \cite{zerospartfunfermi1,zerospartfunfermi2}) systems are studied in many papers. Difficulty in direct experimental observation of
these zeros relates to impossibility of preparing a many-body system with complex parameters. However, in articles \cite{zerospartfuncspin2,ylesdfehfmd}
this problem was solved. The authors for the first time made experimental research on the detection of the density function of zeros on the Lee-Yang circles.
This study is based on the analysis of isothermal magnetization of the Ising ferromagnet FeCl$_2$ in an axial magnetic field. In papers \cite{zerospartfuncspin3,zerospartfuncspin4}
it was suggested to measure the Lee-Yang zeros in the time domain. The relation of the Lee-Yang zeros of the Ising ferromagnet with the decoherence of the probe spin was obtained.
So, direct experimental observation of Lee-Yang zeros on trimethylphosphite molecule was reported in \cite{zerospartfuncspin1}. The methods
of the detection of dynamical Lee-Yang zeros was considered in \cite{eddlyz}. In \cite{zerospartfunbose4} the possibility of experimental observation
of Lee-Yang zeros of an interacting Bose gas was proposed. The relation between zeros of two-time correlation function of probe spin and zeros of partition
function of spin bath was found in \cite{zerospartfuncspin7}. This relation gives a new possibility for experimental detection of Lee-Yang zeros.

In the present paper we find the relation between the Lee-Yang zeros of the ferromagnetic high-spin bath and observed values,
such as magnetization, susceptibility and higher derivatives of magnetization, of probe spin.
In the third section we present our results for some models, namely, triangle spin cluster, Ising model with
long-range interaction and 1D Ising model with nearest-neighbor interaction. Also we propose
the experimental implementation of these consideration on real physical systems. Conclusions are presented in the last section.

\section{Connection between the Lee-Yang zeros and observation values of probe spin}

We consider the system of $N$ spins $s$ described by a general Ising model with ferromagnetic interaction under a magnetic field $h$.
The Hamiltonian of this system has form (\ref{bathham}), where the spin operators correspond to the spins $s$ and their projections on some direction
take the values $-s\leq m\leq s$. Using the Lee-Yang theorem for the case of an arbitrary high-spin bath \cite{genforarbspin1,genforarbspin2,genforarbspin3}
the partition function of this system at temperature $T$ can be expressed as an $2sN$th polynomial of $z\equiv\exp{\left(-\beta h\right)}$ as (\ref{partfunc2}).

Let us assume that spin bath defined by Hamiltonian $H_{b}$ (\ref{bathham}) being in thermodynamic equilibrium. Then we use a probe spin-$s_0$ coupled
to the bath. The general Hamiltonian of the system takes the form
\begin{eqnarray}
H=H_{b}+H_{p}+H_{i},
\label{generalham}
\end{eqnarray}
where $H_{p}=-h_0s_0^z$ is the Hamiltonian of the probe spin which interacts with the magnetic field of the value $h_0$, and $H_{i}=\lambda s_0^z\sum_{i}s_i^{z}$
describes the interaction between the bath and the probe spins with the coupling constant $\lambda$. It is worth noting that the results which we obtain
for the Ising interaction between spins of bath are valid for the case of isotropic Heisenberg interaction between these spins.

The evolution of system (\ref{generalham}) can be represent as follows
\begin{eqnarray}
\rho(t)=e^{-iHt}\rho(0)e^{iHt}=e^{-iH_pt}e^{-iH_it}\rho(0)e^{iH_it}e^{iH_pt},
\label{evolutionin}
\end{eqnarray}
where the initial state has the form $\rho(0)=\vert\Psi(0)\rangle\langle\Psi(0)\vert e^{-\beta H_{b}}/Z\left(\beta, h\right)$.
Here $\vert\psi(0)\rangle=\sum_{m=-s_0}^{s_0}a_m\vert m\rangle$ is the initial state of the probe spin, with normalization condition
$\sum_{m=-s_0}^{s_0}\vert a_m \vert^2=1$, which is determined by the complex parameters $a_m$ and is spanned by the basis vectors $\vert m \rangle$.
The basis vectors are the eigenstates of the $z$-components of spin operator with value $m$. We use the system of units where the Plank constant $\hbar=1$.
To obtain eq. (\ref{evolutionin}) we use the fact that $H_{b}$, $H_{p}$ and $H_{i}$ mutually commute which allows us to get rid of $\exp{\left(-iH_{b}t\right)}$
part of the evolution operator. The evolution of the system due to the interaction between the probe and bath spins can be expressed as follows
\begin{eqnarray}
e^{-iH_{i}}\rho(0)e^{iH_{i}}=\sum_{m,k=-s_0}^{s_0}a_ma_k^*e^{-i\lambda (m-k)t\sum_{i}s_i^z}\frac{e^{-\beta H_b}}{Z\left(\beta, h\right)}\vert m\rangle\langle k\vert.\nonumber
\end{eqnarray}
So, using the above results the evolution of system (\ref{generalham}) takes the form
\begin{eqnarray}
\rho(t)=\sum_{m,k=-s_0}^{s_0}a_ma_k^*e^{ih_0(m-k)t}e^{-i\lambda (m-k)t\sum_{i}s_i^z}\frac{e^{-\beta H_b}}{Z\left(\beta, h\right)}\vert m\rangle\langle k\vert,
\label{evolution}
\end{eqnarray}

We consider the evolution of the probe spin by averaging expression (\ref{evolution}) over the states of the remaining system
\begin{eqnarray}
\rho_p(t)=\sum_{m,k=-s_0}^{s_0}a_ma_k^*e^{ih_0(m-k)t}\frac{Z\left(\beta,h-i\lambda (m-k)t/\beta\right)}{Z\left(\beta, h\right)}\vert m\rangle\langle k\vert,
\label{probeevol}
\end{eqnarray}
where we use the fact that $\textrm{Tr}\left[e^{-i\lambda (m-k)t\sum_{i}s_i^z-\beta H_b}\right]=Z\left(\beta,h-i\lambda (m-k)t/\beta\right)$.
This expression contains the partition functions of the bath system with a complex magnetic fields $h-i\lambda (m-k)t/\beta$ for different values of $m-k$.
Each of these functions vanish when the time of evolution $t$ is such that $\exp{\left(-\beta h+i\lambda (m-k)t\right)}$ equals to Lee-Yang zeros.
So, we can observe the Lee-Yang zeros of bath system by exploration of the evolution of probe system.

Examining the evolution of the probe spin under the action of the bath we can observe the Lee-Yang zeros of this bath. For this purpose,
we calculate the mean value of the operator $(s_0^+)^{m-k}$ as a function of time
\begin{eqnarray}
&&\langle\left(s_0^+\right)^{m-k}\rangle = \textrm{Tr}\left[\rho_p(t)\left(s_0^+\right)^{m-k}\right]\nonumber\\
&&=\frac{Z\left(\beta,h+i\lambda (m-k)t/\beta\right)}{Z\left(\beta, h\right)}e^{-ih_0(m-k)t}\nonumber\\
&&\times \sum_{l=-s_0}^{s_0-(m-k)}a_la_{l+m-k}^*\prod_{q=1}^{m-k}\sqrt{s_0(s_0+1)-(l+q-1)(l+q)},\nonumber\\
\label{averagevalue}
\end{eqnarray}
where $s_0^+=s_0^x+is_0^y$ is the ladder operator. Using representation (\ref{partfunc2}) and taking $h=h_0=0$ we can rewrite this expression as follows
\begin{eqnarray}
&&\langle\left(s_0^+\right)^{m-k}\rangle = \frac{e^{iNs\lambda(m-k)t}\prod_{n=1}^{2Ns}\left(e^{-i\lambda(m-k)t}-e^{i\theta_n}\right)}{\prod_{n=1}^{2Ns}\left(1-e^{i\theta_n}\right)}\nonumber\\
&&\times \sum_{l=-s_0}^{s_0-(m-k)}a_la_{l+m-k}^*\nonumber\\
&&\times\prod_{q=1}^{m-k}\sqrt{s_0(s_0+1)-(l+q-1)(l+q)}.
\label{averagevalue3}
\end{eqnarray}
As we can see, for the specific Lee-Yang zero $z_n$ the following condition $-\lambda (m-k)t_n=\theta_n$ is satisfied.
The values of $t_n=-\theta_n/\left(\lambda (m-k)\right)$ which correspond to zeros of expression
(\ref{averagevalue3}) are related to the Lee-Yang zeros of partition function of the bath. Also we can see that the greater $m-k$, the faster this expression
vanishes. Therefore, to detect the zeros of the partition function of bath it is enough to measure mean value (\ref{averagevalue3}) for particular value
of $m-k$. It is easy to verify that a similar connection (\ref{averagevalue3}) exists in the case of isotropic Heisenberg interaction
between the spins of the bath.

Measurement the mean value of $(s_0^+)^{m-k}$ as a function of time allows us to observe a Lee-Yang zeros.
So, if $m-k=1$ then to observe the Lee-Yang zeros the $x$ and $y$ components
of magnetization should be measured. If $m-k=2$ then the components of the magnetic susceptibility should be  measured for observation of the Lee-Yang zeros.
For larger values of $m-k$ it is necessary to measure the higher derivatives of magnetization for this purpose. However, as we mentioned earlier, to observe
the Lee-Yang zeros of the partition function of the bath it is enough to measure the mean value (\ref{averagevalue3}) with particular value of $m-k$.
Therefore, further we will present our results for $m-k=1$ $\left(\langle s_0^+\rangle =\langle s_0^x\rangle+i\langle s_0^y\rangle\right)$.
The experimental techniques for measurement of magnetization as a funtion of time is described in supplementary materials of paper \cite{zerospartfuncspin1}.
Let us study this problem in detail in the case of different models of the bath.

\section{Models of bath and their application}

In this section we apply our results to some models. Namely, we examine the connection between the Lee-Yang zeros of the spins bath with different
structures and observation values of probe spin. Also we suggest the physical realization of these considerations.

\subsection{Triangle spin cluster}

\begin{figure}[!!h]
\center{\includegraphics[scale=0.35, angle=0.0, clip]{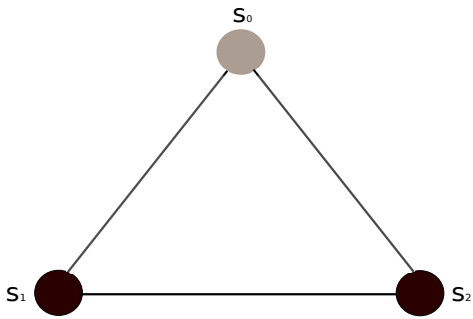}}
\caption{Triangle spin cluster. The model consists of two spins $s_1$ and $s_2$ (the black circles) as the bath and spin $s_0$ (the gray circle) as the probe.}
\label{trianglesc1}
\end{figure}

\begin{figure}[!!h]
\includegraphics[scale=0.17, angle=0.0, clip]{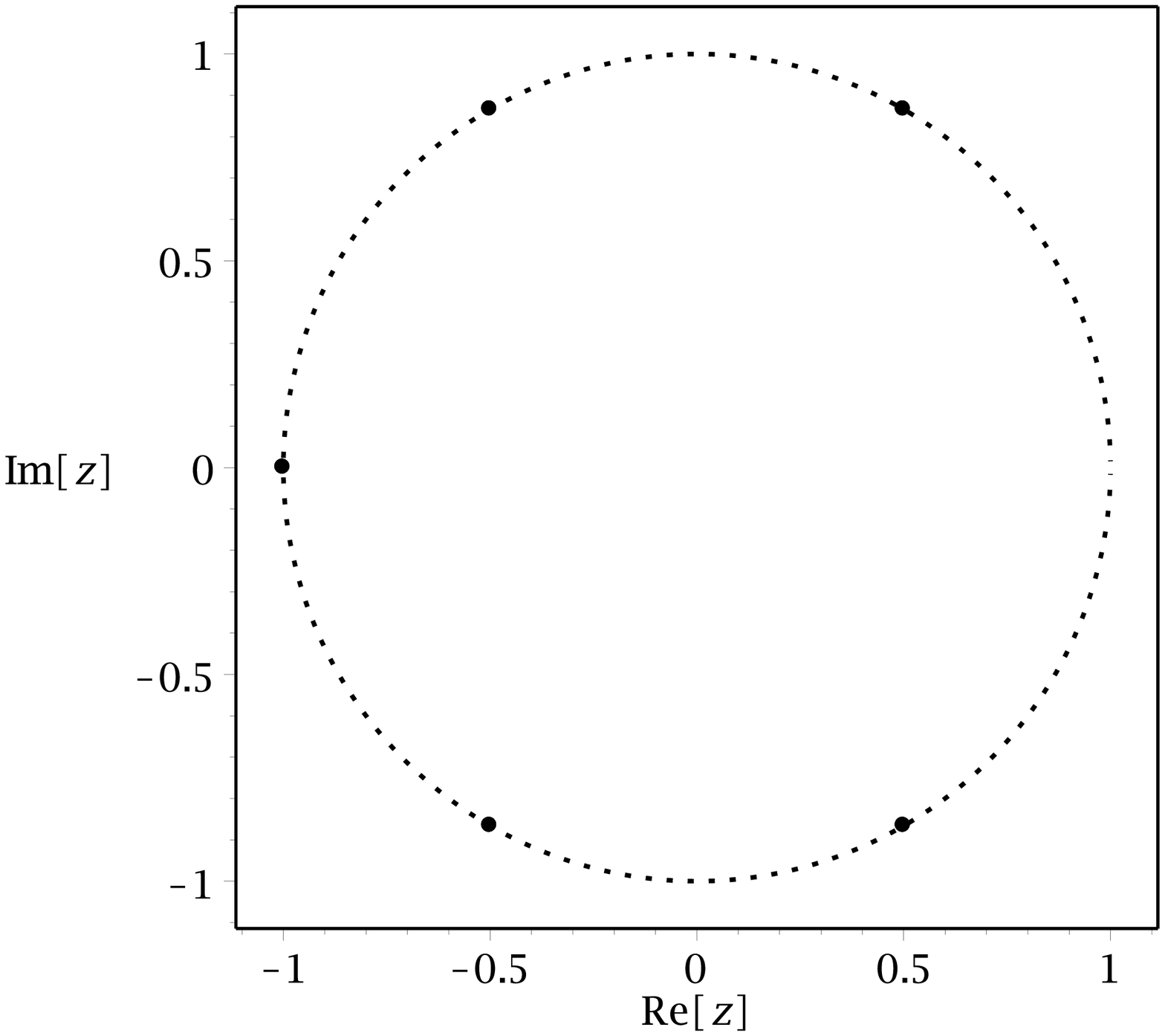}
\includegraphics[scale=0.17, angle=0.0, clip]{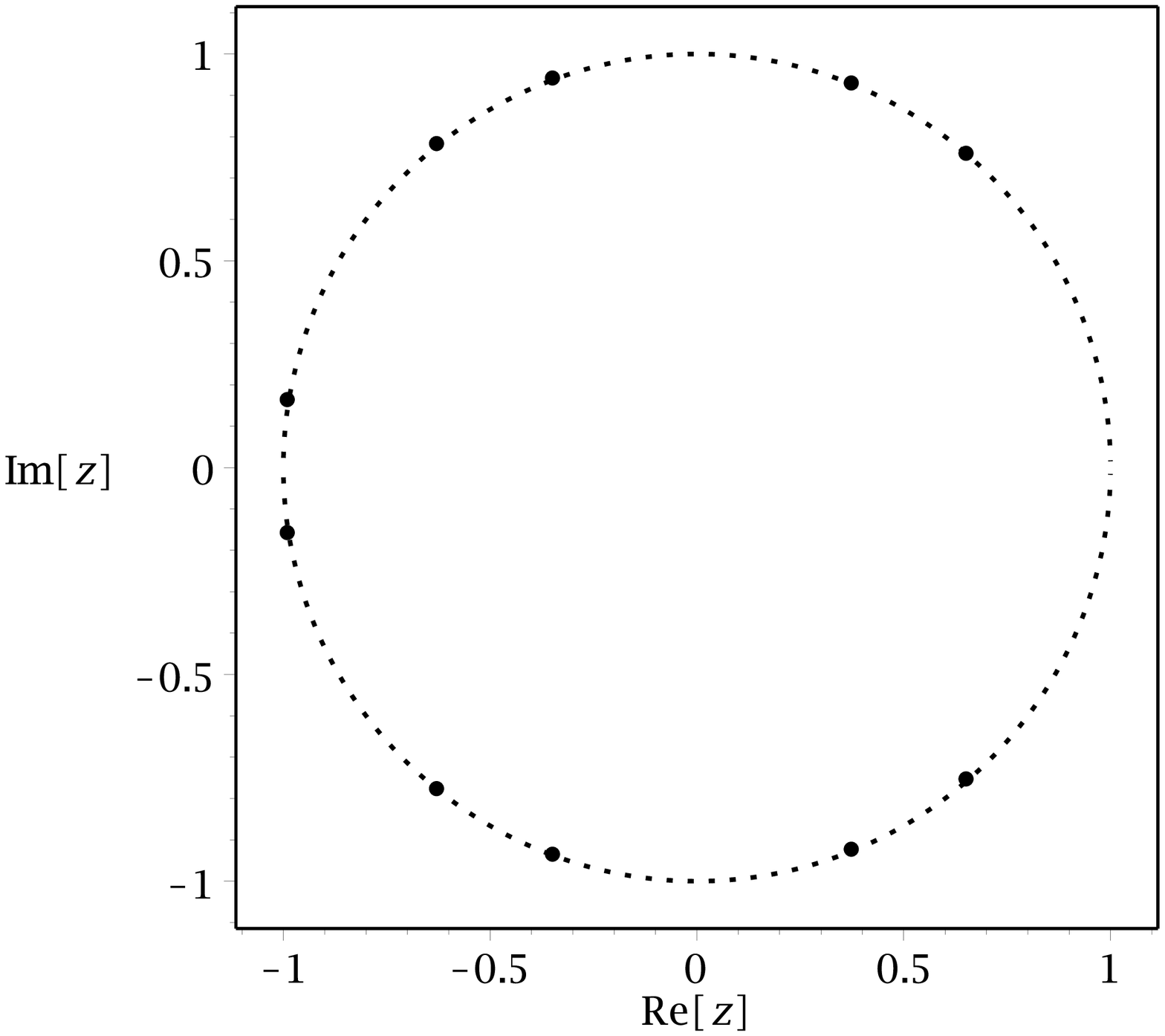}
\includegraphics[scale=0.17, angle=0.0, clip]{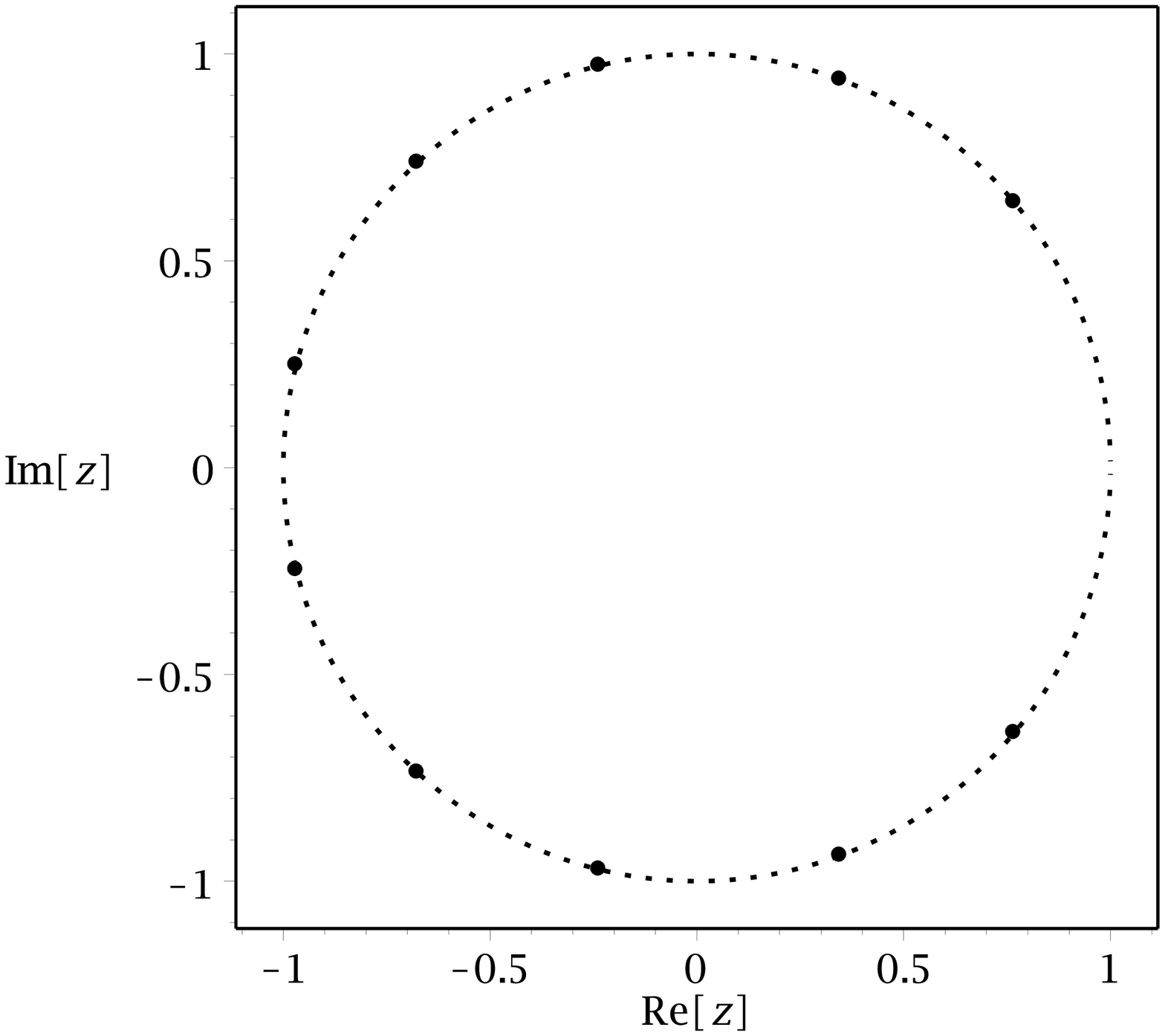}
\includegraphics[scale=0.17, angle=0.0, clip]{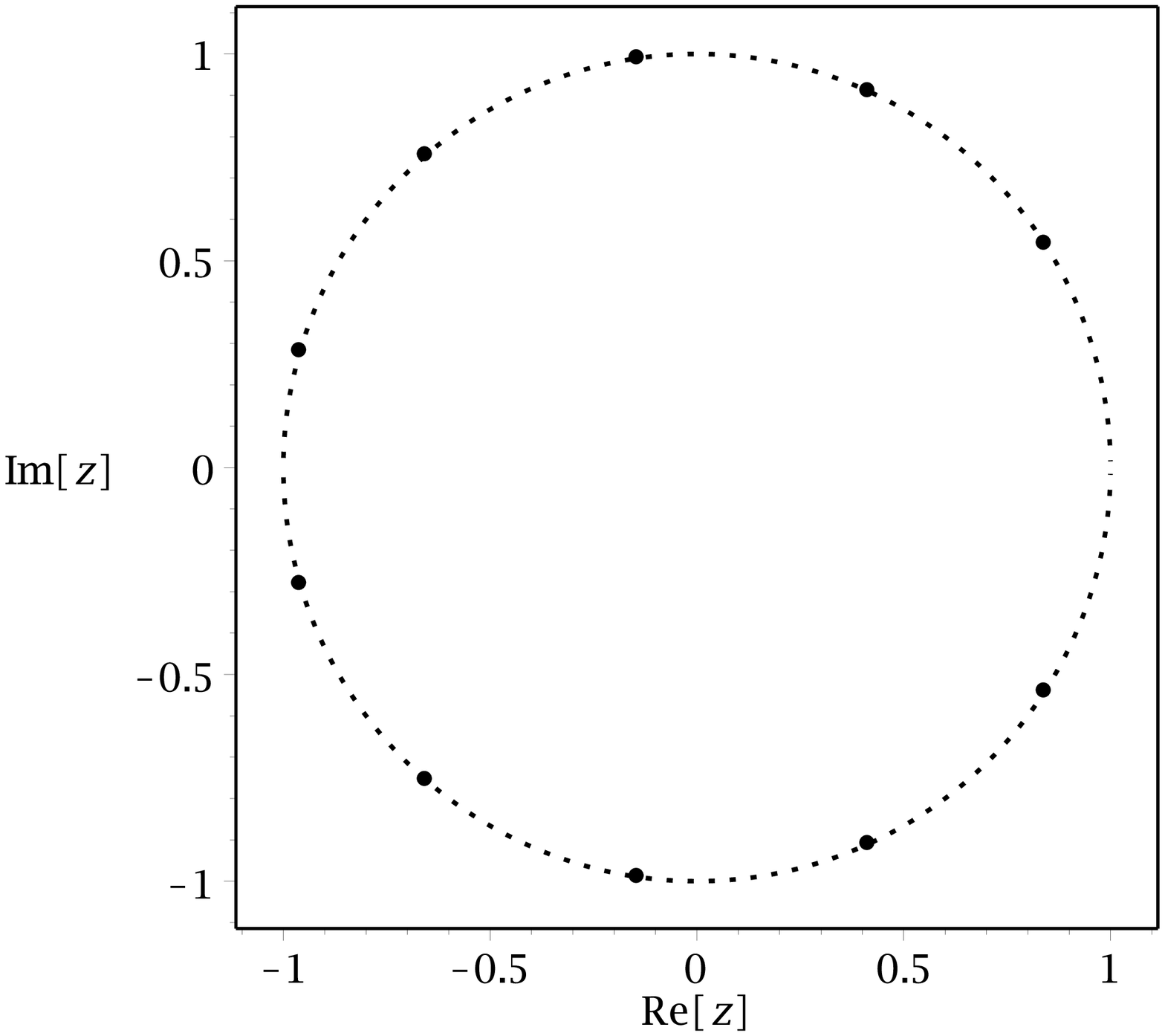}\\
\subcaptionbox{\label{}}{\includegraphics[scale=0.16, angle=0.0, clip]{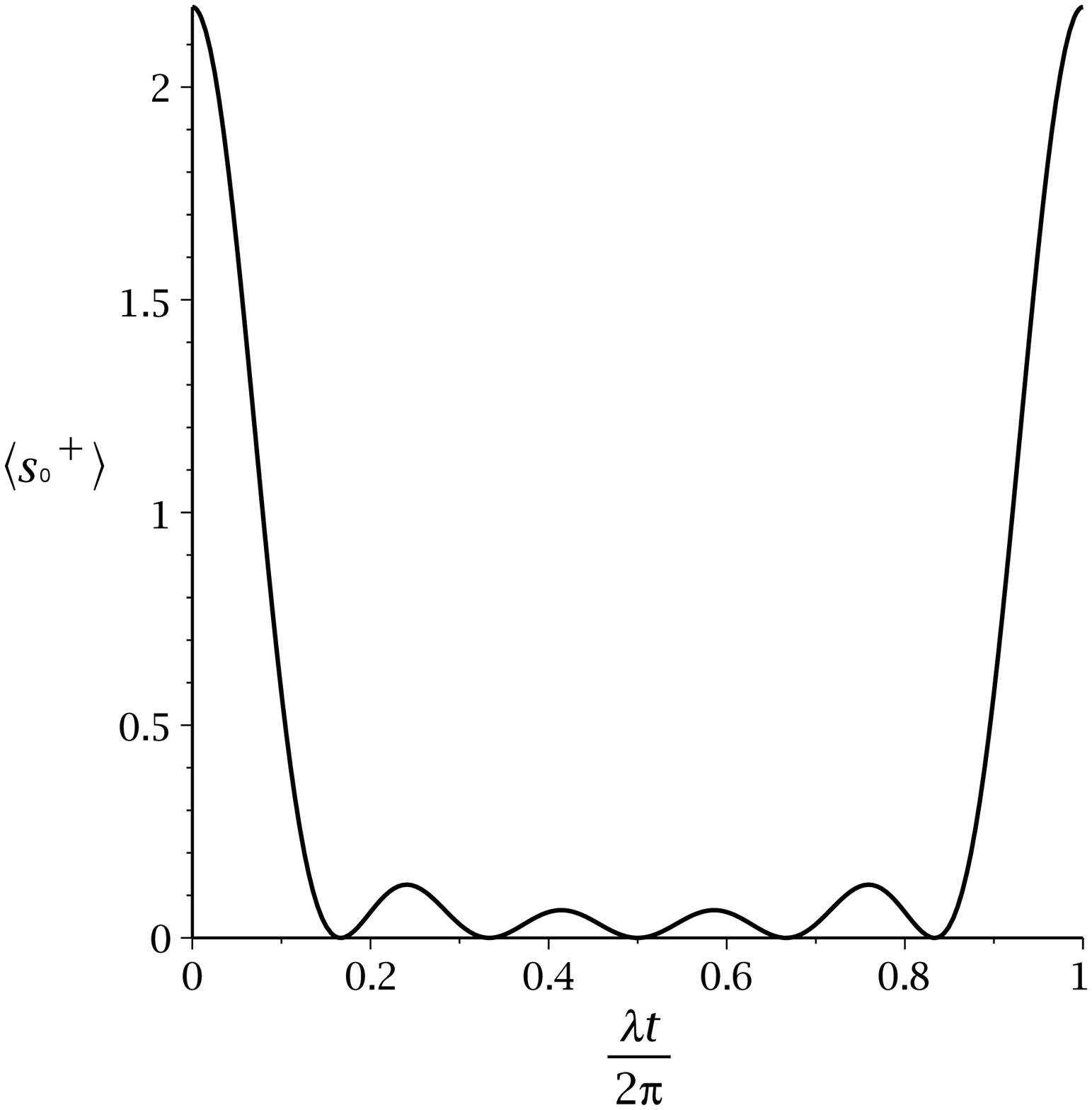}}
\subcaptionbox{\label{}}{\includegraphics[scale=0.16, angle=0.0, clip]{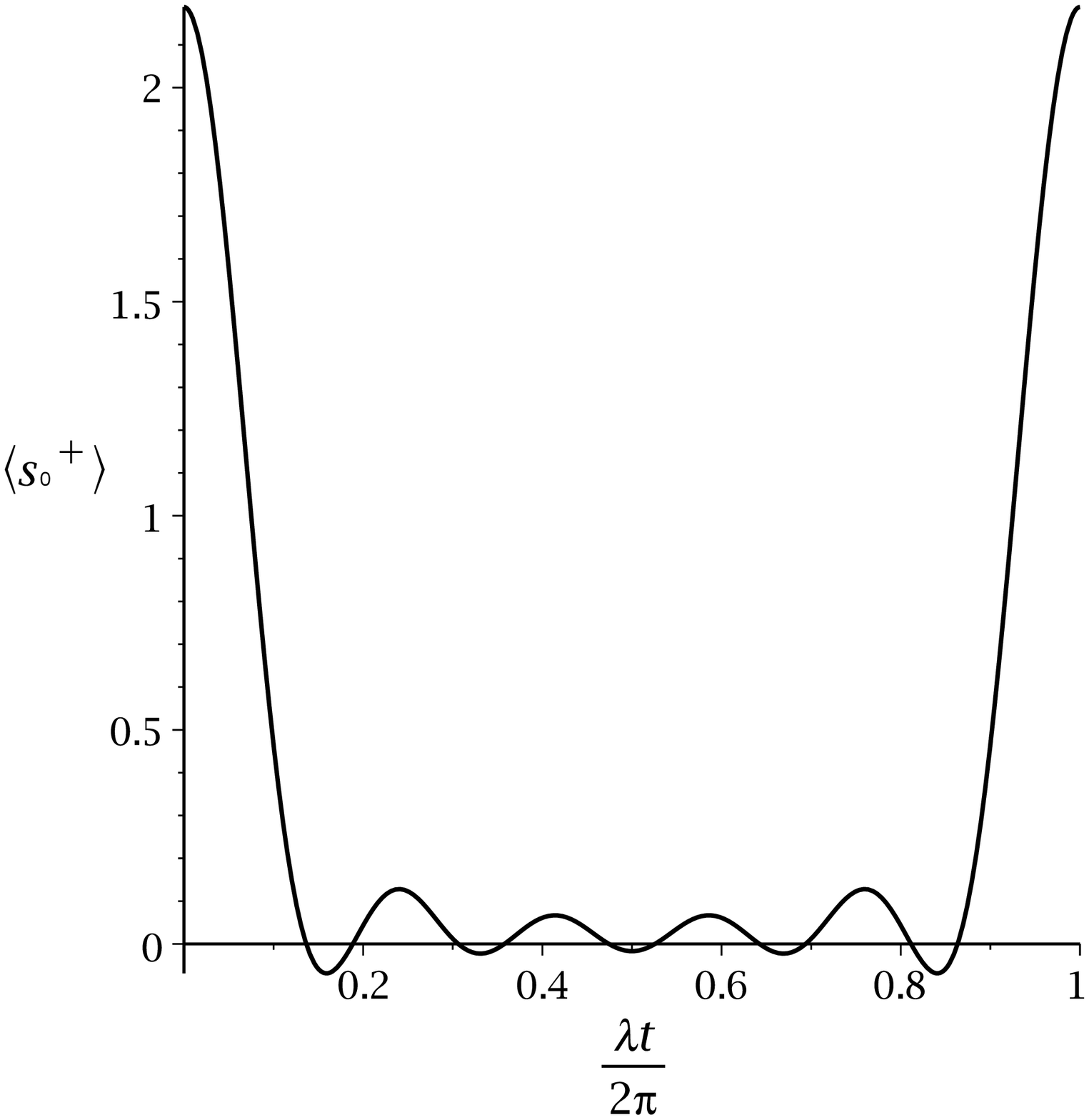}}
\subcaptionbox{\label{}}{\includegraphics[scale=0.16, angle=0.0, clip]{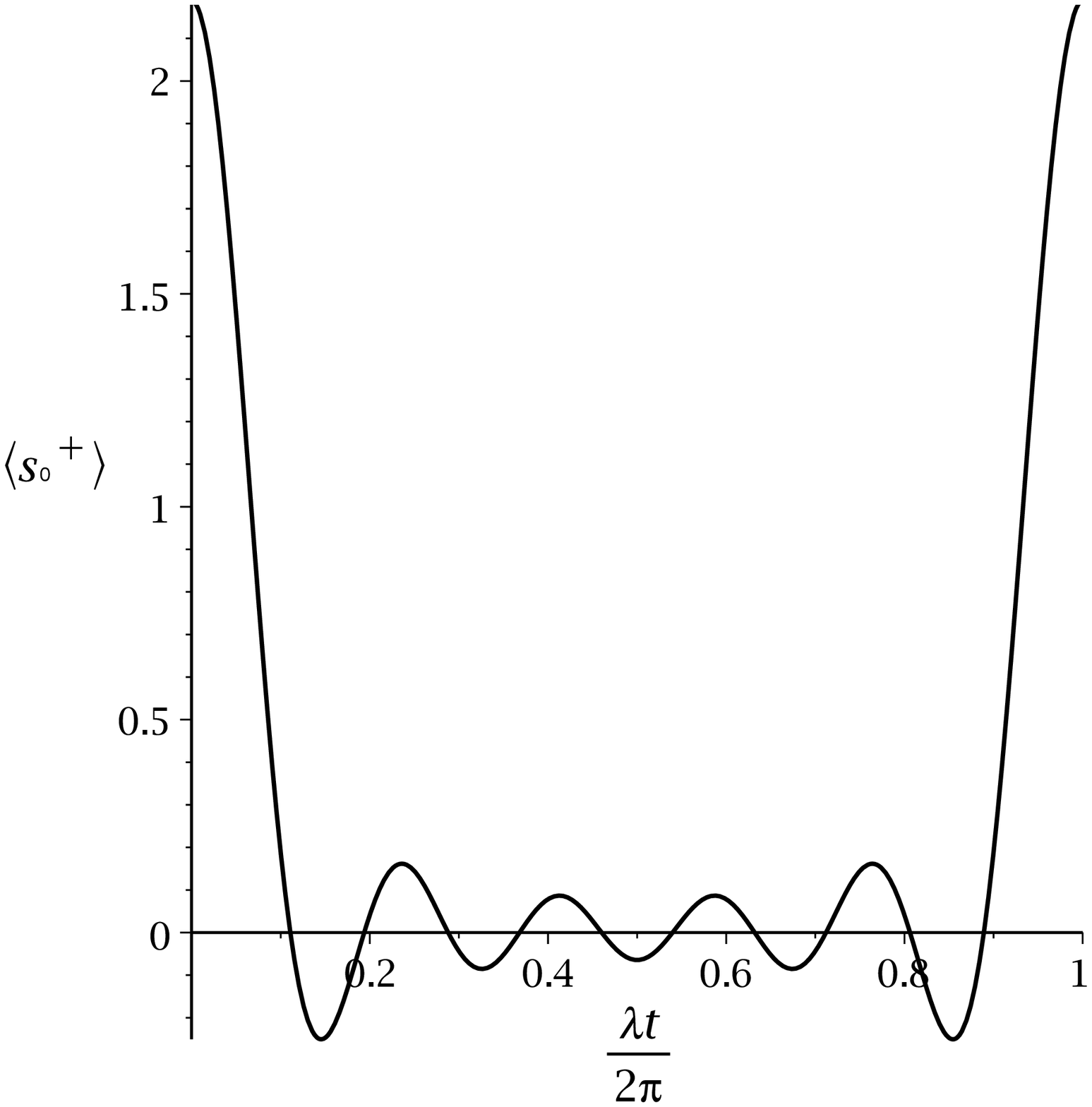}}
\subcaptionbox{\label{}}{\includegraphics[scale=0.16, angle=0.0, clip]{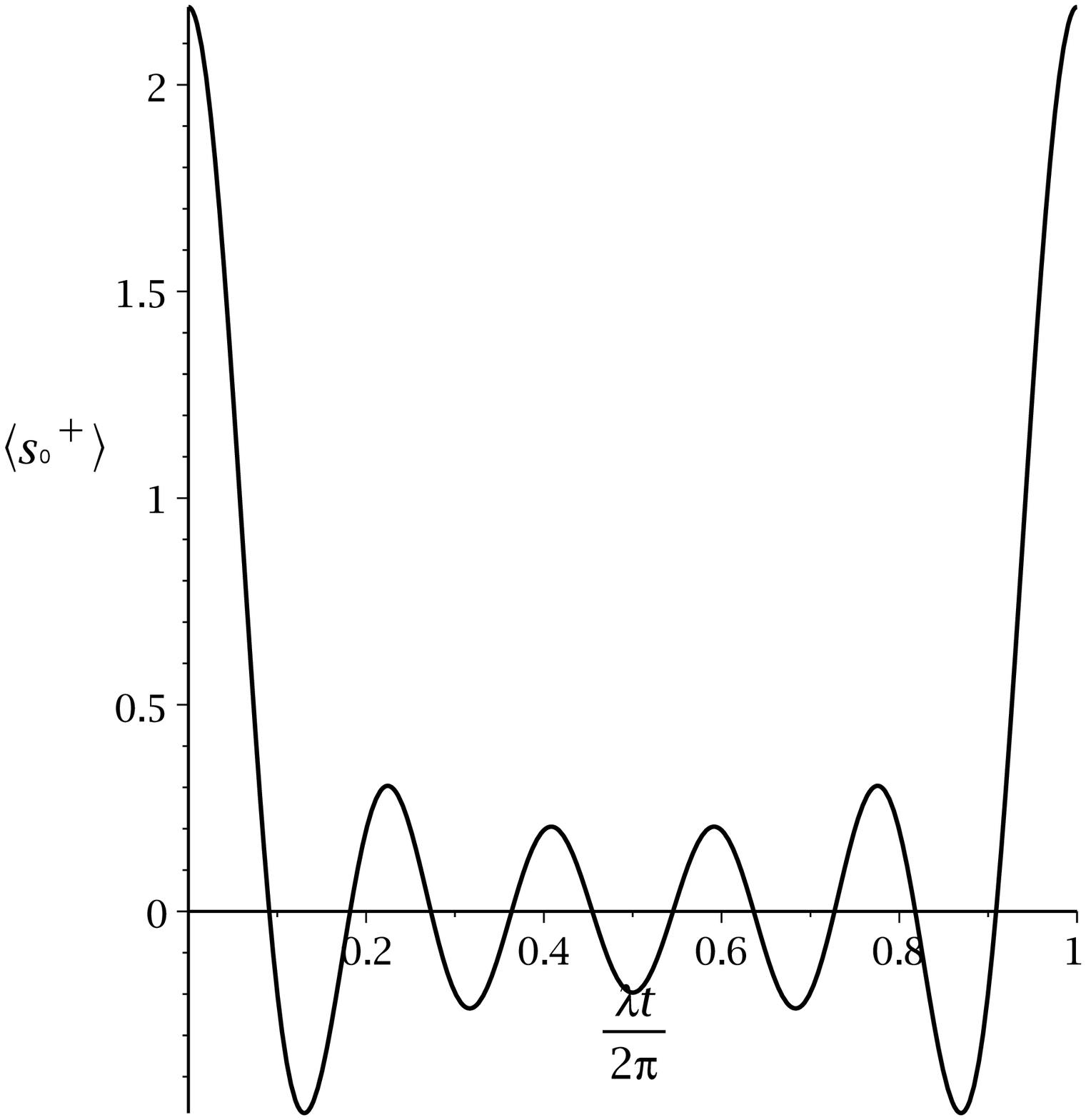}}
\caption{Correspondence between the Lee-Yang zeros of partition function (\ref{trianglepartf}) of bath composed of two spin-$5/2$ and mean value of
the $s_0^x$ operator (\ref{averagevalue}) with $h=h_0=0$ of the probe spin-$5/2$ as a function of time. The results are presented for different temperatures: (a) $T=\infty$, (b) $T=32J$, (c)
$T=8J$ and (d) $T=J$.}
\label{LYZTSC}
\end{figure}

First of all, let us study the triangle spin cluster which consists of two spins $s_1$ and $s_2$ as the bath and one spin $s_0$ as the probe spin
(see fig. \ref{trianglesc1}). The interaction between bath spins is ferromagnetic and is described by Ising Hamiltonian. Also
the interaction between the spin bath and the probe spin is defined by Ising Hamiltonian. The Hamiltonian of the complete system has the form
\begin{eqnarray}
H=-Js_1^z s_2^z+\lambda s_0^z\left(s_1^z+s_2^z\right),
\label{triangleham}
\end{eqnarray}
The connection between the Lee-Yang zeros of the bath and observable values of the probe spin is determined by eq. (\ref{averagevalue})
with $h=h_0=0$. Here the partition function of two-spin bath in the external magnetic field $h$ has the following form
\begin{eqnarray}
Z\left(\beta,h\right)=\sum_{m_1=-s_1}^{s_1}\sum_{m_2=s_2}^{s_2}e^{\beta Jm_1m_2}e^{\beta h (m_1+m_2)},
\label{trianglepartf}
\end{eqnarray}
where $s_1$, $s_2$ are the values of each spin of the bath and $m_1$, $m_2$ their projection on the $z$-axis, respectively.

Let us consider the obtained results on a real physical system of manganese ferrite (MnFe$_2$O$_4$) \cite{mnferrite1,mnferrite2,mnferrite3}.
The manganese ferrite contains two ions of Fe$^{3+}$ as a bath spins and one ion of Mn$^{2+}$ as a probe spin. Each ion has spin $5/2$.
The interaction between spins has the exchange nature and described by isotropic Heisenberg model.
The exchange integrals of this system are calculated in papers \cite{mnferriteexi1,mnferriteexi2,mnferriteexi3}.
Similarly as in paper \cite{zerospartfuncspin1}, the interaction between spins can be simulated by the Ising model (\ref{triangleham}). Indeed,
calculating the partition function at high temperatures (or small $\beta$) the Heisenberg model can be approximated by the Ising model. Fig. \ref{LYZTSC}
shows the Lee-Yang zeros of two-spin bath composed of $5/2$ spins and mean value of $s_0^x$ operator of the probe spin-$5/2$ for different temperatures.
Here and further in the article we take the eigenstate of $s_0^x$ with the highest eigenvalue as an initial state of the probe spin.
This state can be easily prepared in the experiment. Due to the corresponding form of the initial state and the fact that $h=h_0=0$
we obtain that $\langle s_0^y\rangle$ vanishes and $\langle s_0^+\rangle=\langle s_0^x\rangle$. Therefore, on fig. \ref{LYZTSC}
and in further cases we have the time dependence of $\langle s_0^x\rangle$ operator.
Also, it is worth noting that the similar structure and properties have the following ferrites: ZnFe$_2$O$_4$, CoFe$_2$O$_4$, NiFe$_2$O$_4$,
where Zn$^{2+}$, Co$^{2+}$, Ni$^{2+}$ play the roles of the probe spins, respectively.

\subsection{Ising model with long-range interaction}

In this subsection we consider the bath with long-range Ising type interaction. We assume that the interaction couplings between each pair of spins
are the same and equal $J_{ij}=J/N$. Then the partition function takes form (\ref{partfunc}), where
\begin{eqnarray}
&&p_n=e^{\frac{\beta J}{2N}(Ns-n)^2}\nonumber\\
&&\times\frac{1}{2\pi}\int_{0}^{2\pi}e^{-i\phi \left(Ns-n\right)}\left(\sum_{m_1=-s}^se^{-{m_1}^2\frac{\beta J}{2N}}\cos(m_1\phi)\right)^Nd\phi.\nonumber\\
\label{form8}
\end{eqnarray}
Here $m_1$ is the projection of the spin on the $z$-axis. So, for this case the mean value of the probe spin ladder operator is described by eq. (\ref{averagevalue}) without external magnetic field,
where partition function (\ref{partfunc}) has $p_n$ defined by eq. (\ref{form8}). The effective spin system with long-range interaction can be easily prepared using trapped ions \cite{SchrodCat1,EQSSTI,QSDEGHTI} or ultracold atoms
\cite{opticallattice1,opticallattice2,opticallattice3,opticallattice4}.

\begin{figure}
\center{\includegraphics[scale=0.6, angle=0.0, clip]{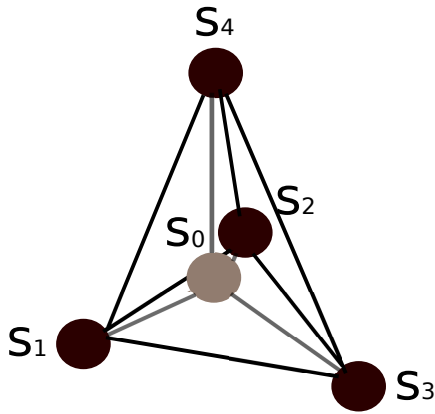}}
\caption{The tetrahedron spin bath $s_i$ (the black circles) interacts with the probe spin $s_0$ (the gray circle). In this model, all the spins
mutually interact. The interaction between the bath spins is ferromagnetic.}
\label{fivespincluster}
\end{figure}

\begin{figure}
\includegraphics[scale=0.17, angle=0.0, clip]{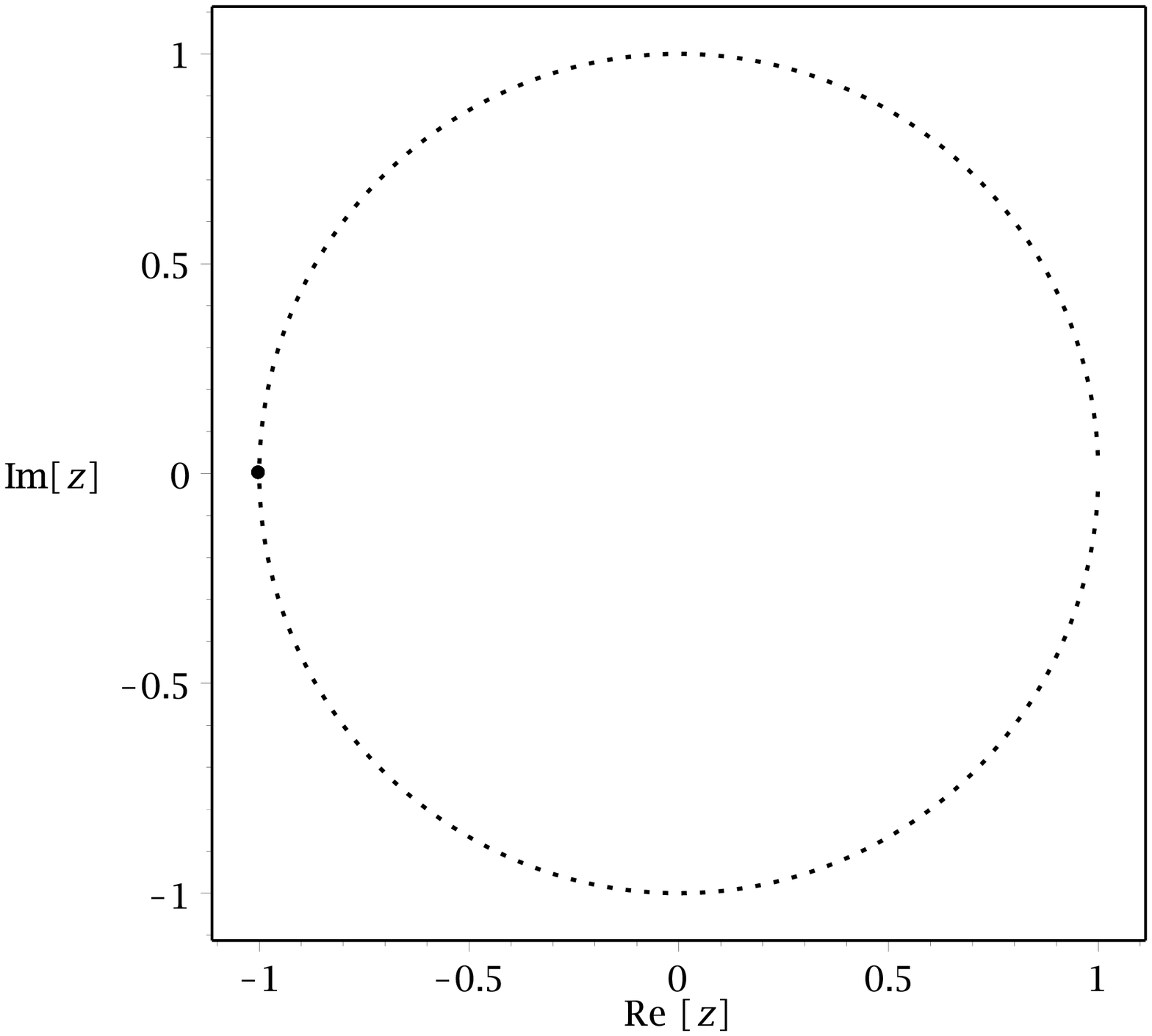}
\includegraphics[scale=0.17, angle=0.0, clip]{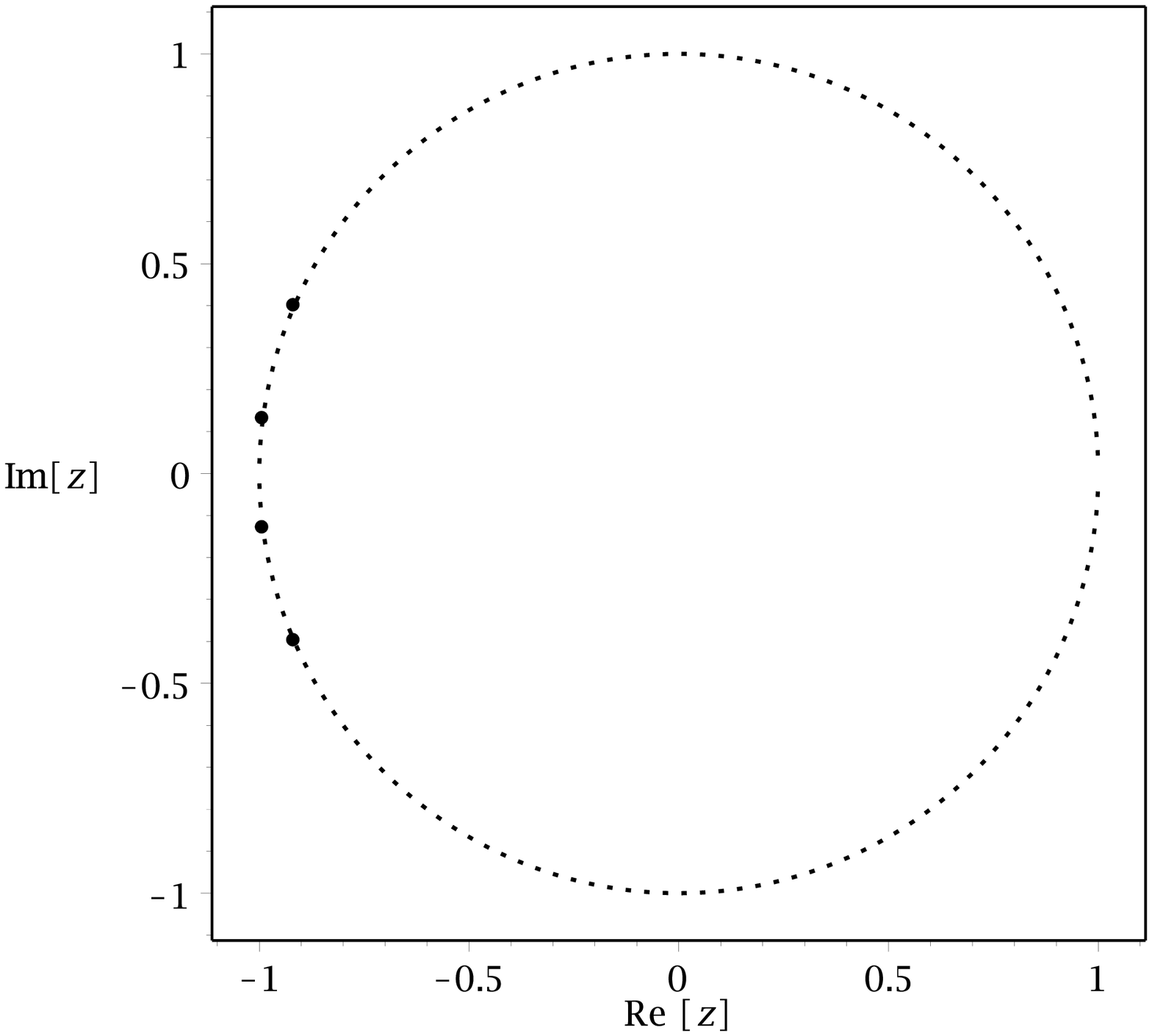}
\includegraphics[scale=0.17, angle=0.0, clip]{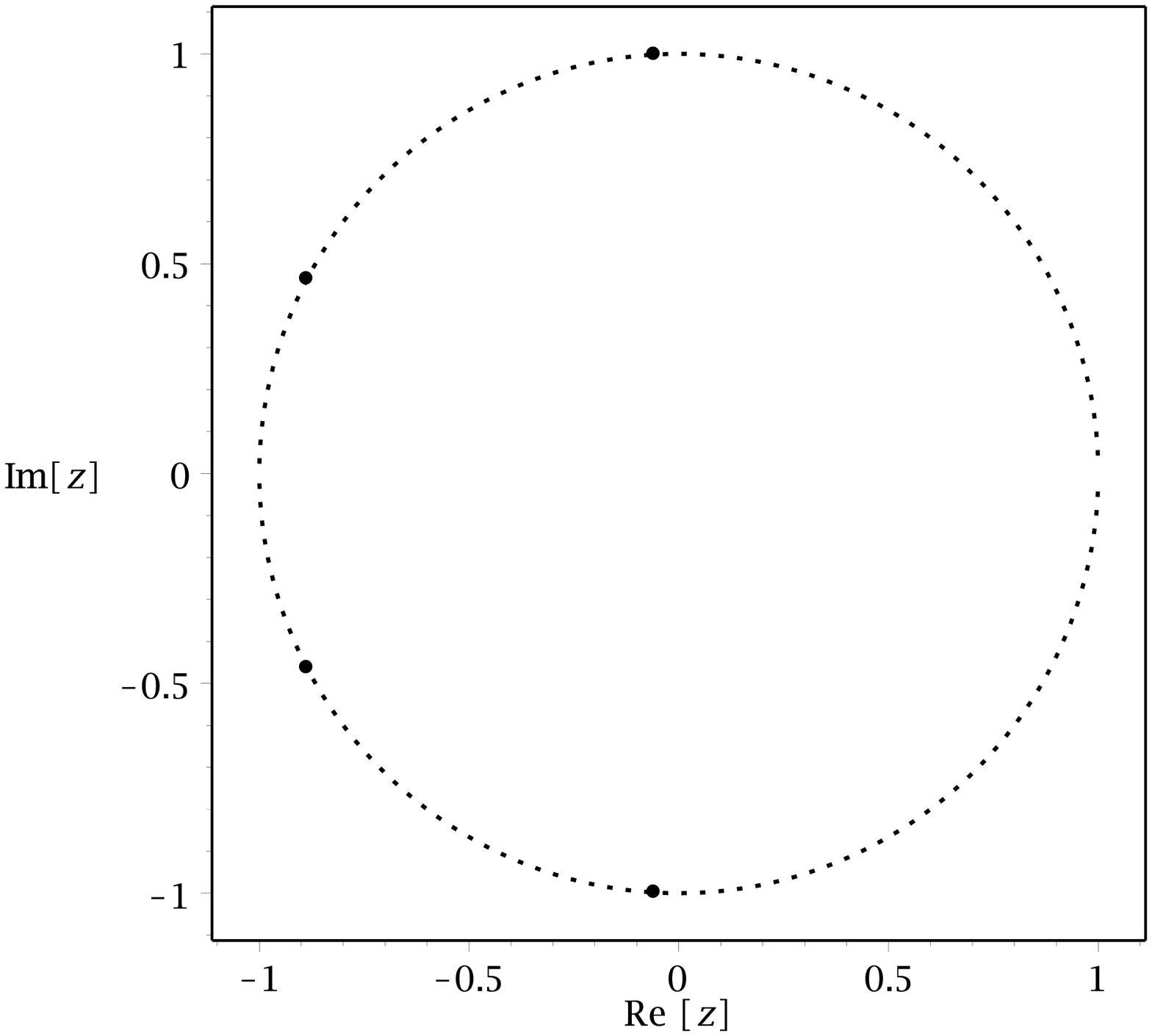}
\includegraphics[scale=0.17, angle=0.0, clip]{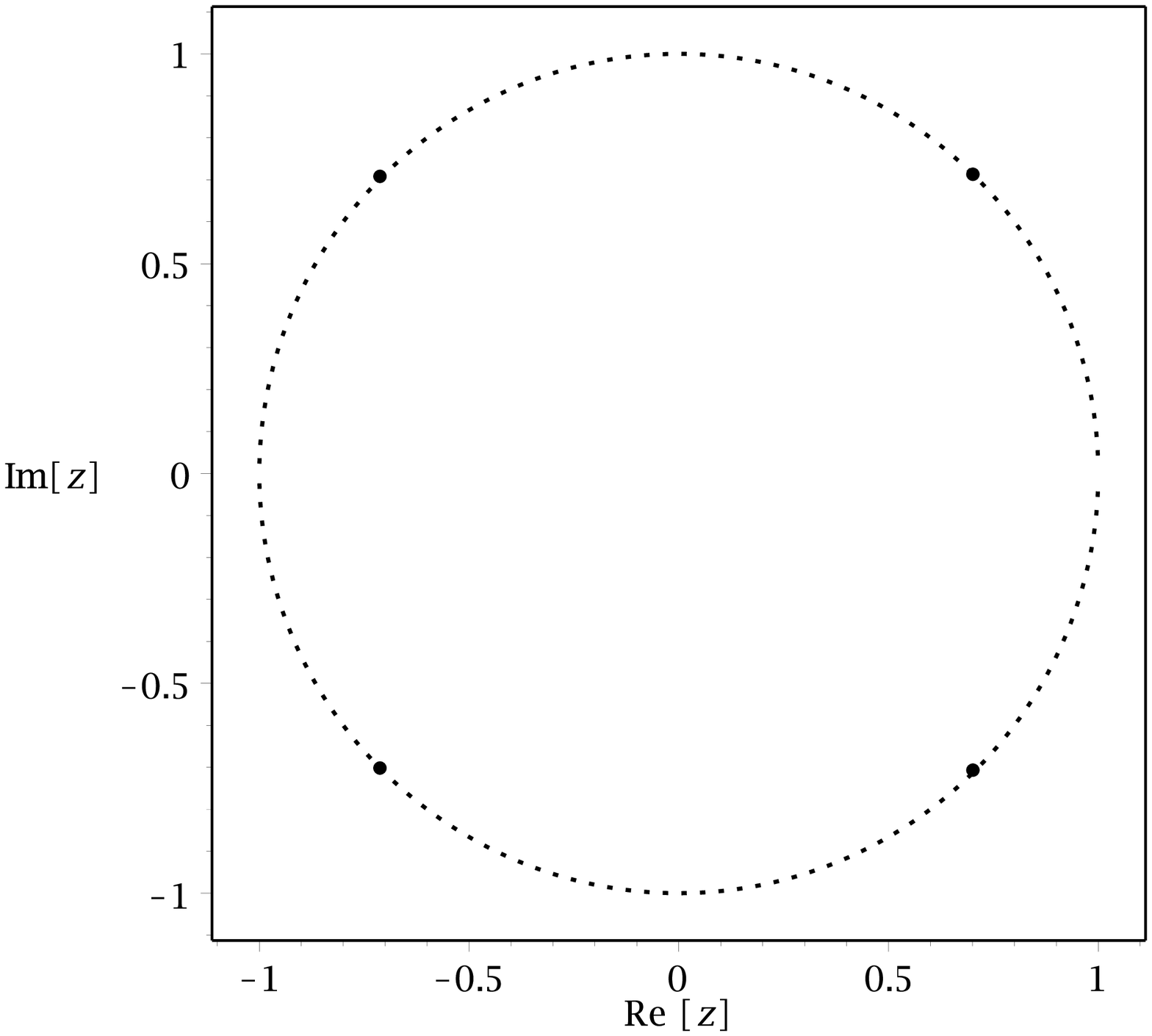}\\
\subcaptionbox{\label{}}{\includegraphics[scale=0.17, angle=0.0, clip]{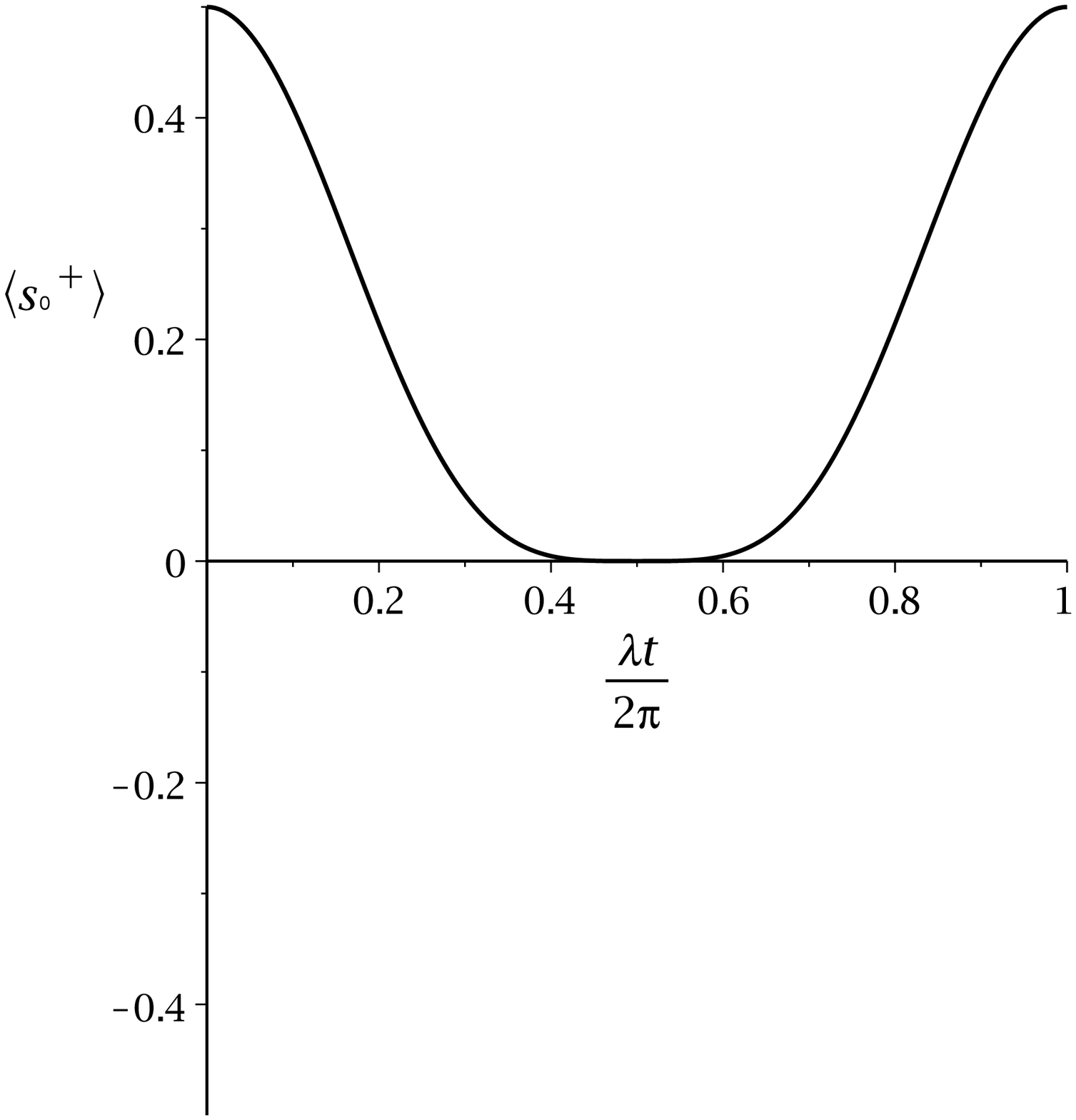}}
\subcaptionbox{\label{}}{\includegraphics[scale=0.17, angle=0.0, clip]{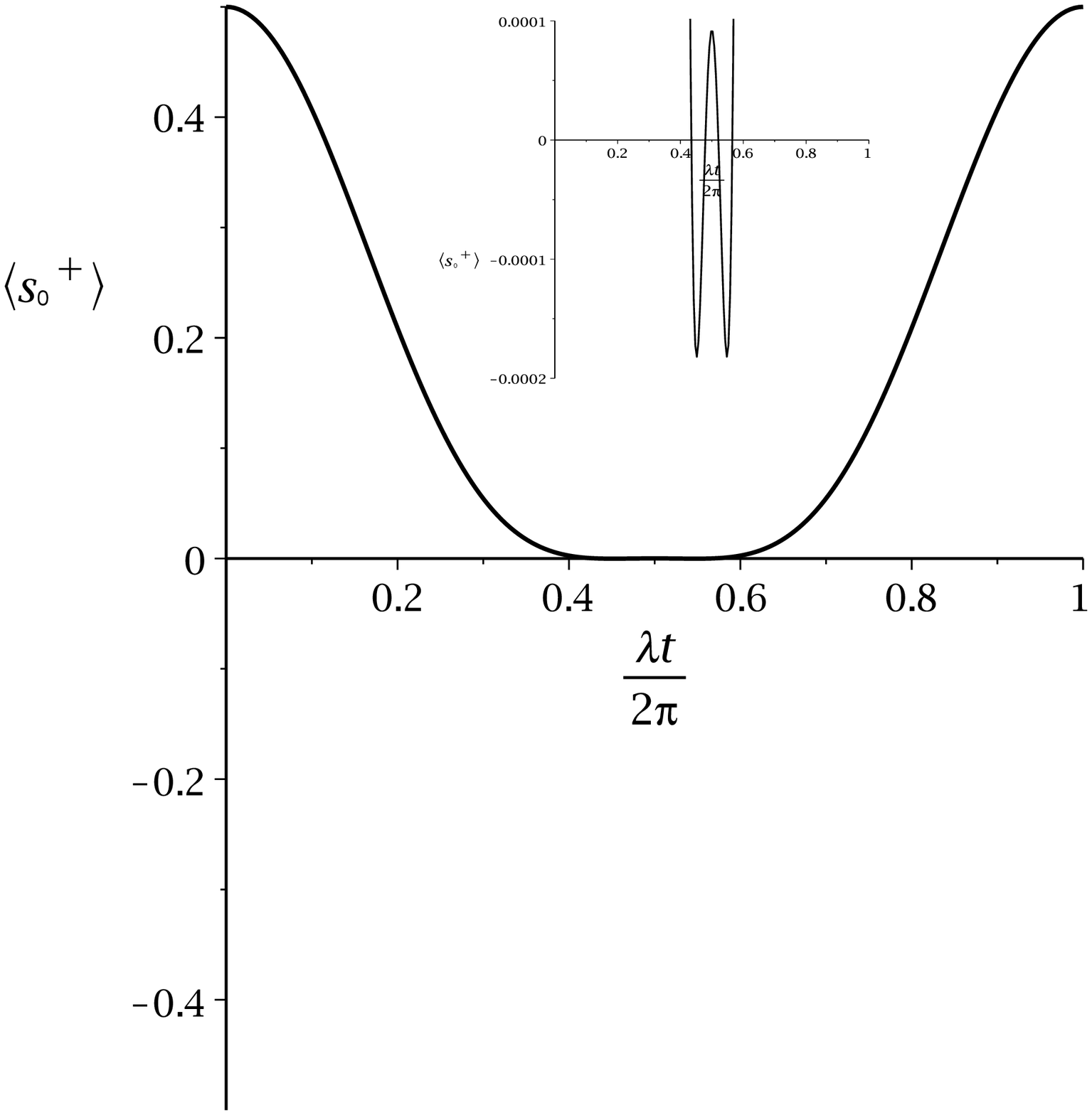}}
\subcaptionbox{\label{}}{\includegraphics[scale=0.17, angle=0.0, clip]{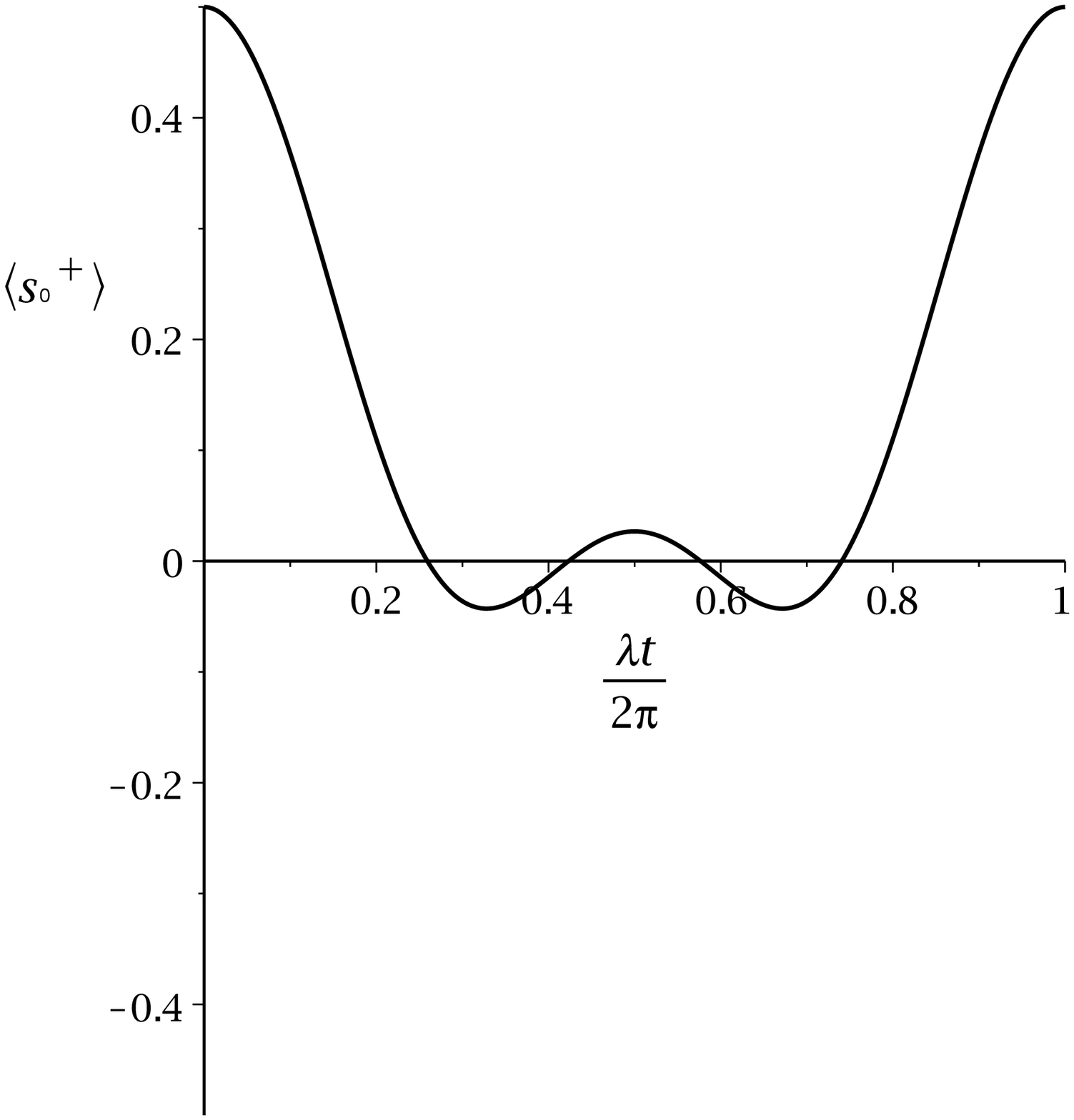}}
\subcaptionbox{\label{}}{\includegraphics[scale=0.17, angle=0.0, clip]{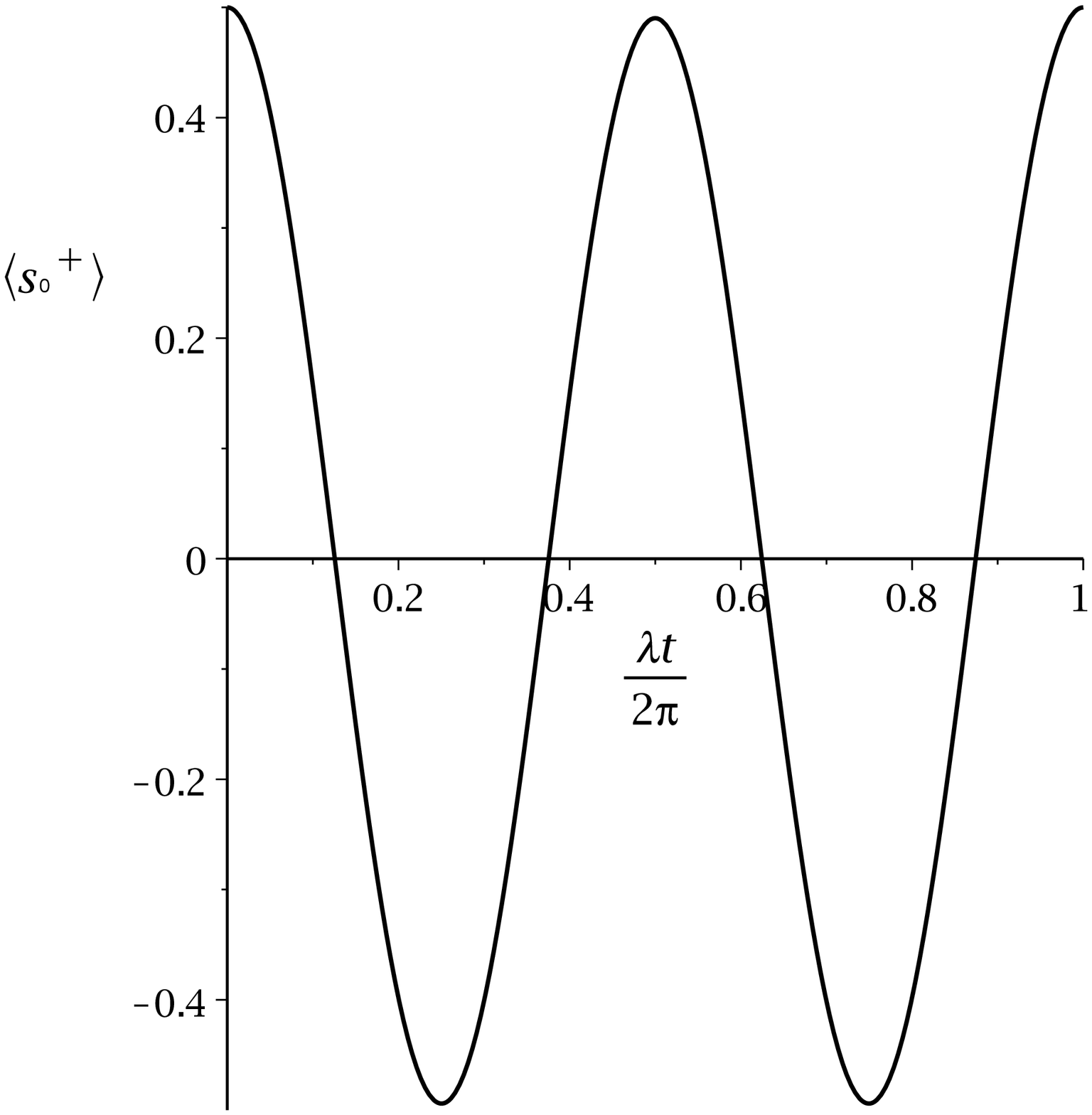}}
\caption{The connection between the Lee-Yang zeros of partition function (\ref{partfunc}) with $p_n$ defined by eq. (\ref{form8})
and mean value of the $s_0^x$ operator of the probe spin-$1/2$ (\ref{averagevalue}) with $h=h_0=0$ for the four spin-$1/2$ bath.
The results are presented for different temperatures: (a) $T=\infty$, (b) $T=8J$, (c) $T=J/2$ and (d) $T=J/16$.}
\label{lritispin1}
\end{figure}

We express the results of this subsection on the five-spin cluster (fig. \ref{fivespincluster}), where the bath composed of four
spins $s$ which form tetrahedron and interact between themselves due to ferromagnetic Ising-type interaction. The interaction with the probe spin-$s_0$
is also described by the Ising model. Such a structure has the methane molecule with $^{13}$C. The molecule CH$_4$ consists of four atoms of $^{1}$H
and one atom of $^{13}$C. Each of them has nuclear spin $1/2$. The interaction coupling between the proton
spins is ferromagnetic and equals $-12.4$ Hz, and the interaction coupling of each proton spin with carbon nucleus spin equals $125$ Hz \cite{fivespincluster}.
We describe the interaction between atom spins in this molecule by the Ising model. In fig. \ref{lritispin1} we express our results for this structure,
where the spins of protons in hydrogen atoms play the role of bath and the spin of carbon nucleus is the probe. 
It is worth noting that molecules of SiH$_4$, GeH$_4$ and SnH$_4$ have similar structure.

\subsection{1D Ising model with nearest-neighbor interaction}

\begin{figure}[!!h]
\center{\includegraphics[scale=0.35, angle=0.0, clip]{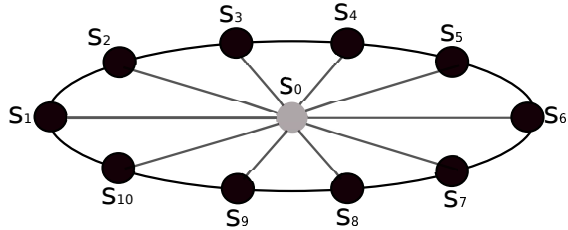}}
\caption{Ring consisting of ten spins $s_i$ (the black circles) as the bath and the probe spin $s_0$ (the gray circle). The interaction between the bath spins
is described by the 1D ferromagnetic Ising model with nearest-neighbor interaction and the interaction of the probe spin with all the spins of the bath
is defined by the Ising model.}
\label{spinringcluster}
\end{figure}

\begin{figure}[!!h]
\includegraphics[scale=0.17, angle=0.0, clip]{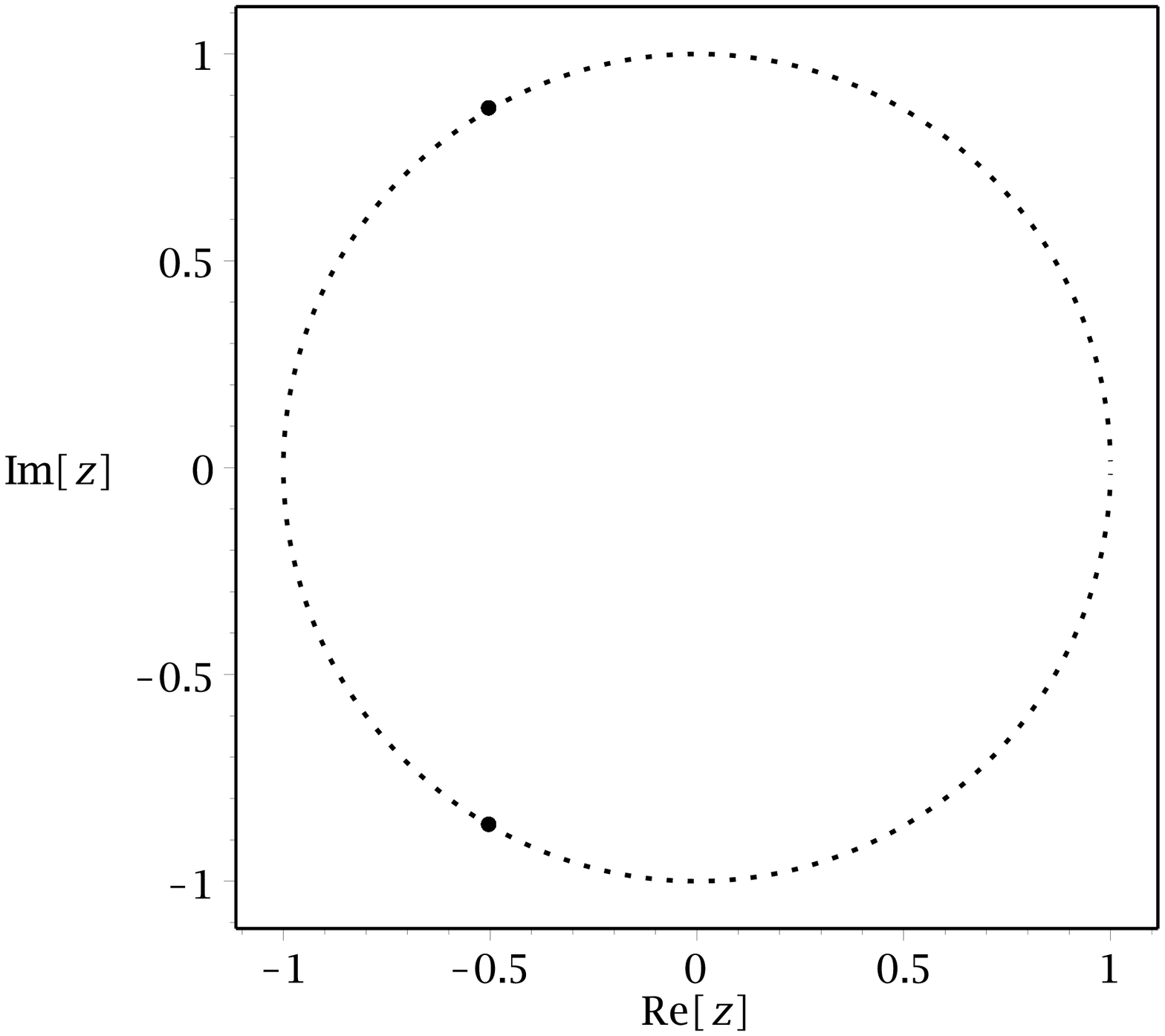}
\includegraphics[scale=0.17, angle=0.0, clip]{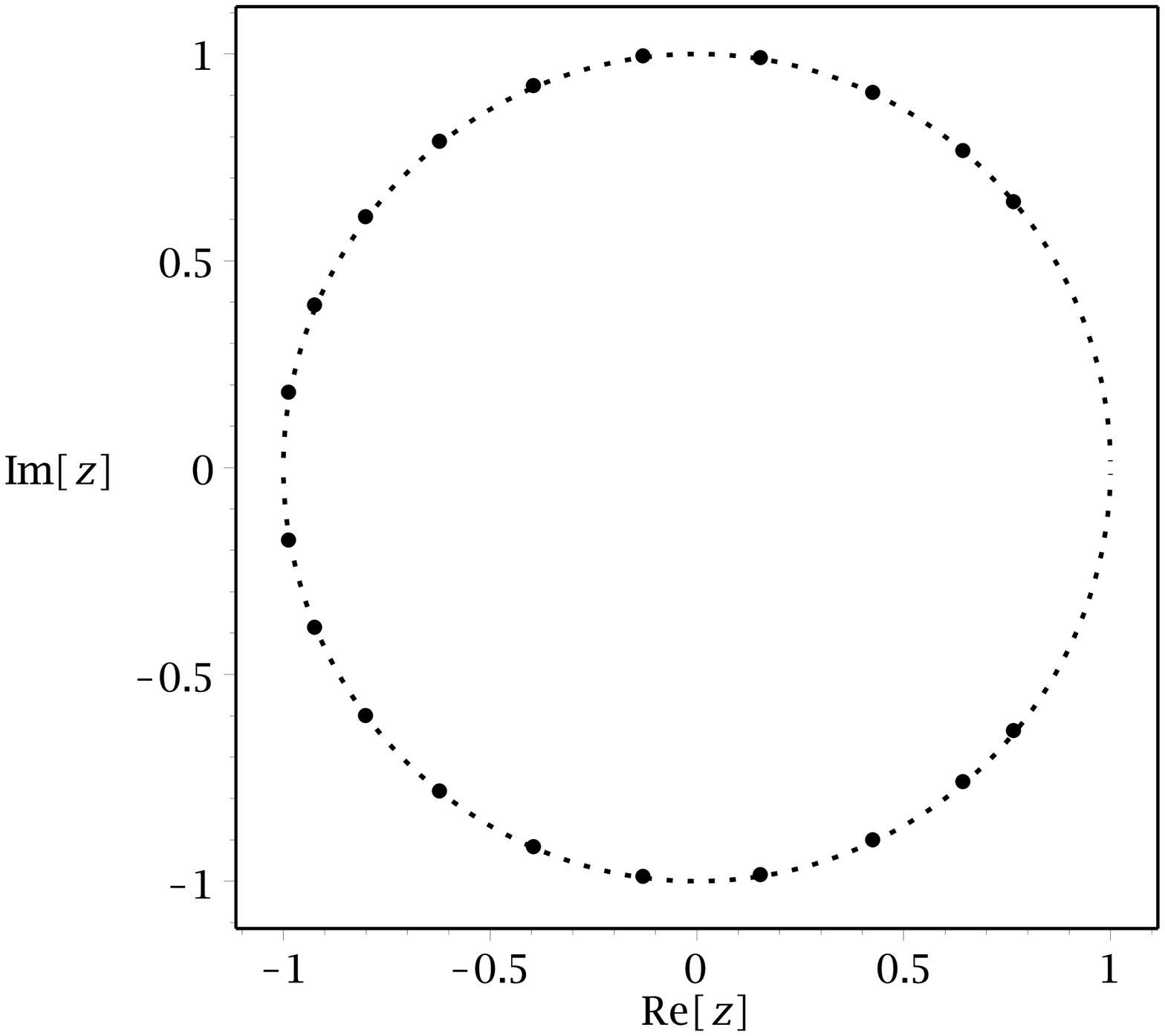}
\includegraphics[scale=0.17, angle=0.0, clip]{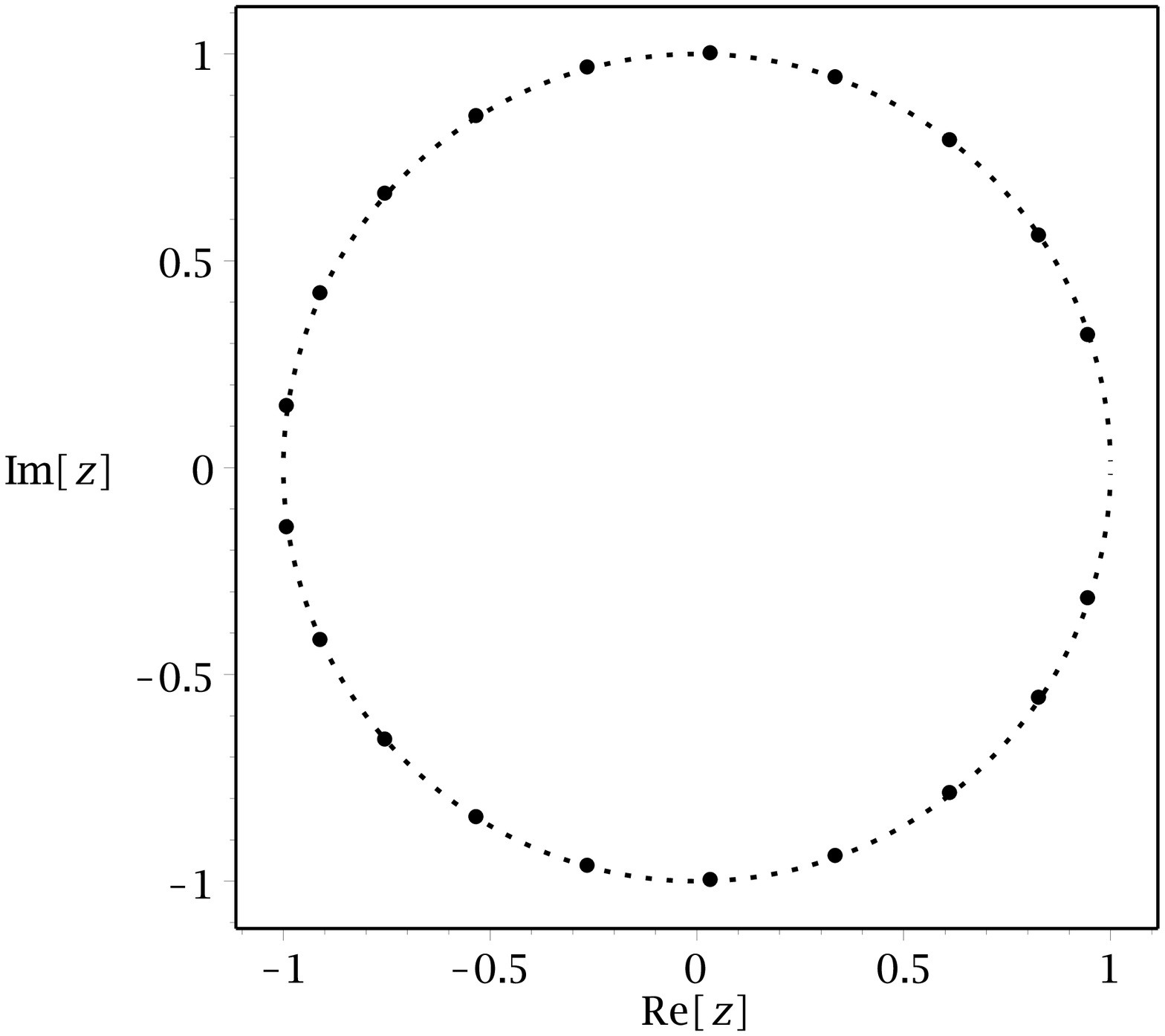}
\includegraphics[scale=0.17, angle=0.0, clip]{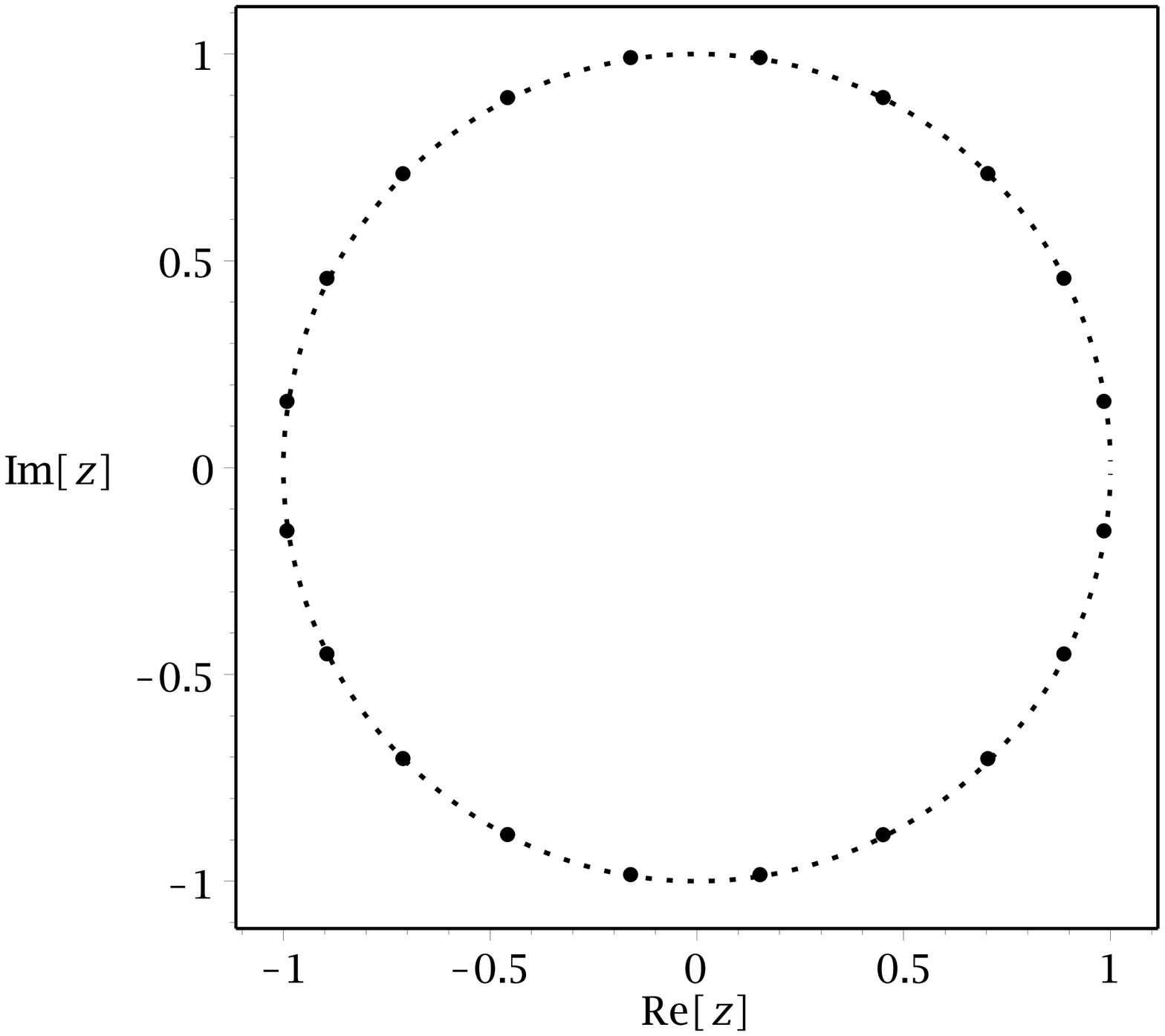}\\
\subcaptionbox{\label{}}{\includegraphics[scale=0.17, angle=0.0, clip]{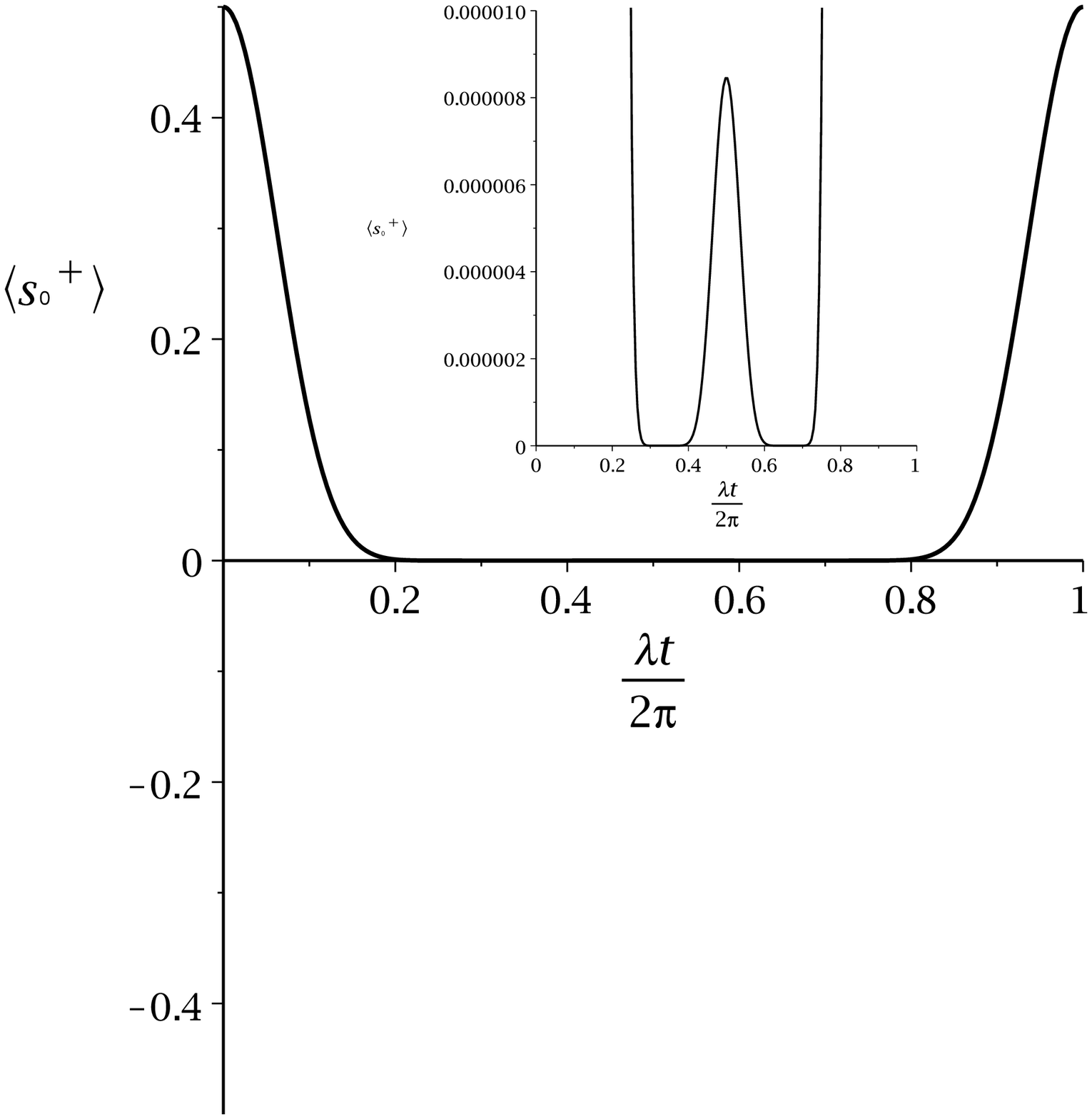}}
\subcaptionbox{\label{}}{\includegraphics[scale=0.17, angle=0.0, clip]{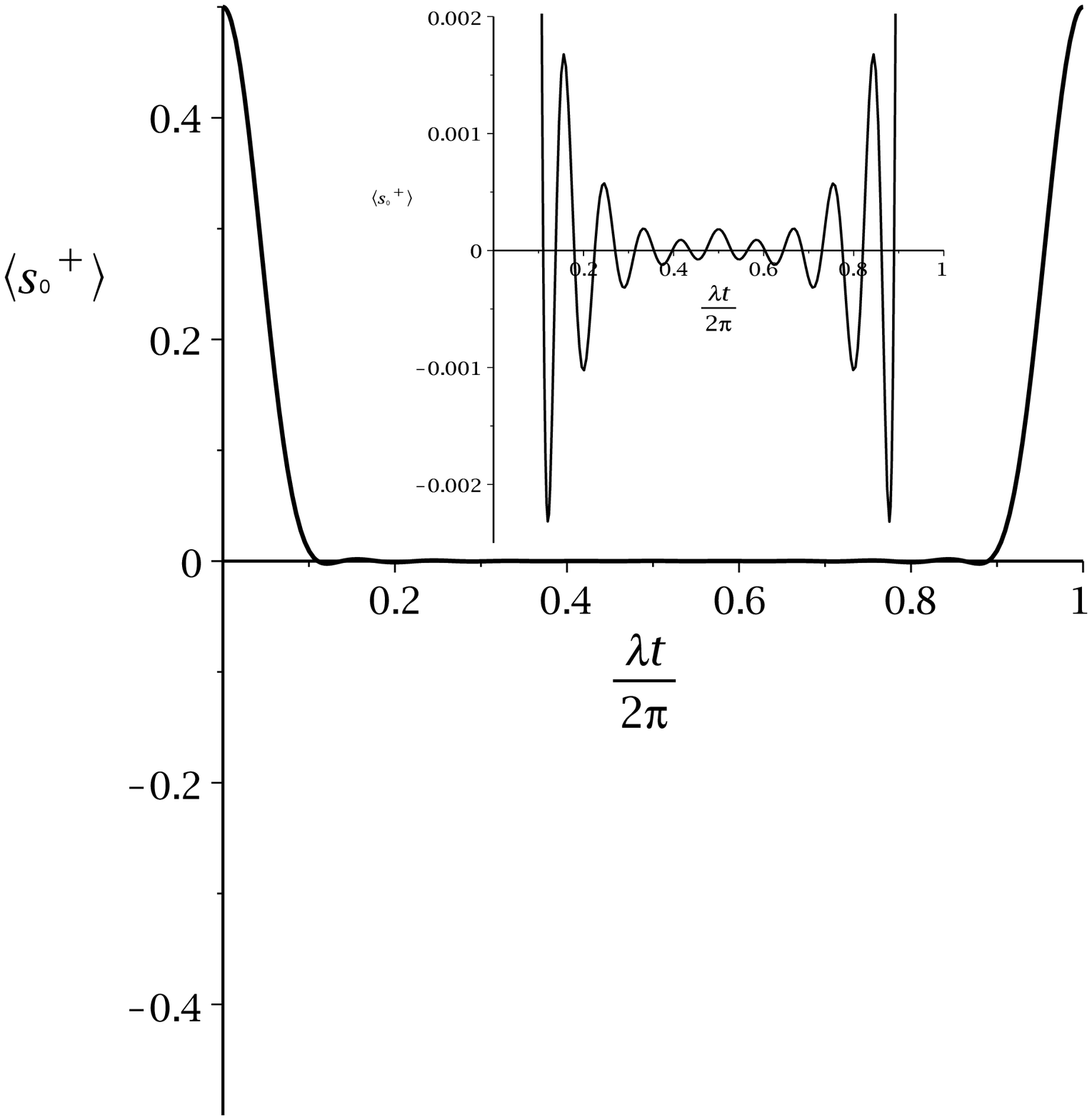}}
\subcaptionbox{\label{}}{\includegraphics[scale=0.17, angle=0.0, clip]{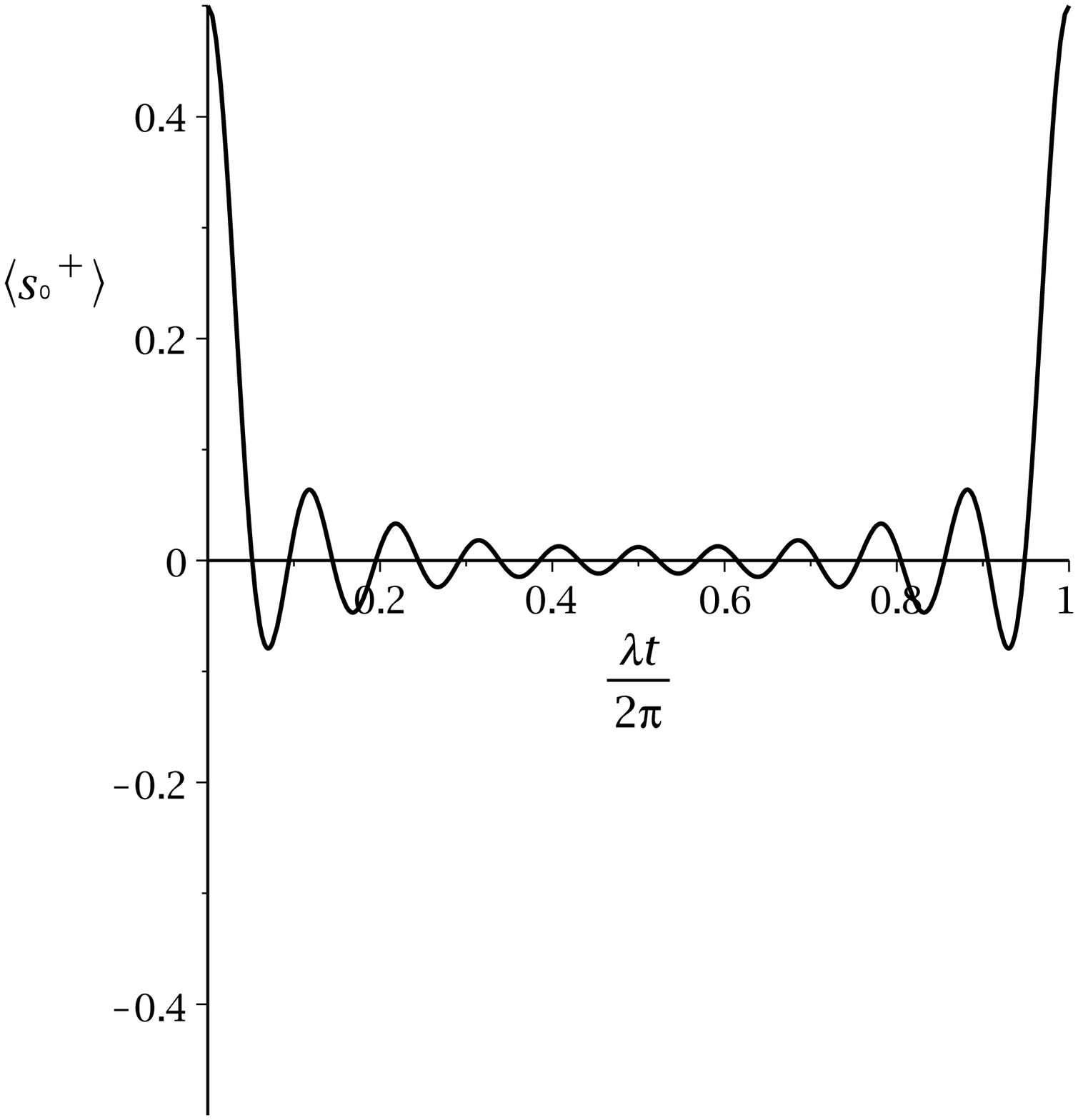}}
\subcaptionbox{\label{}}{\includegraphics[scale=0.17, angle=0.0, clip]{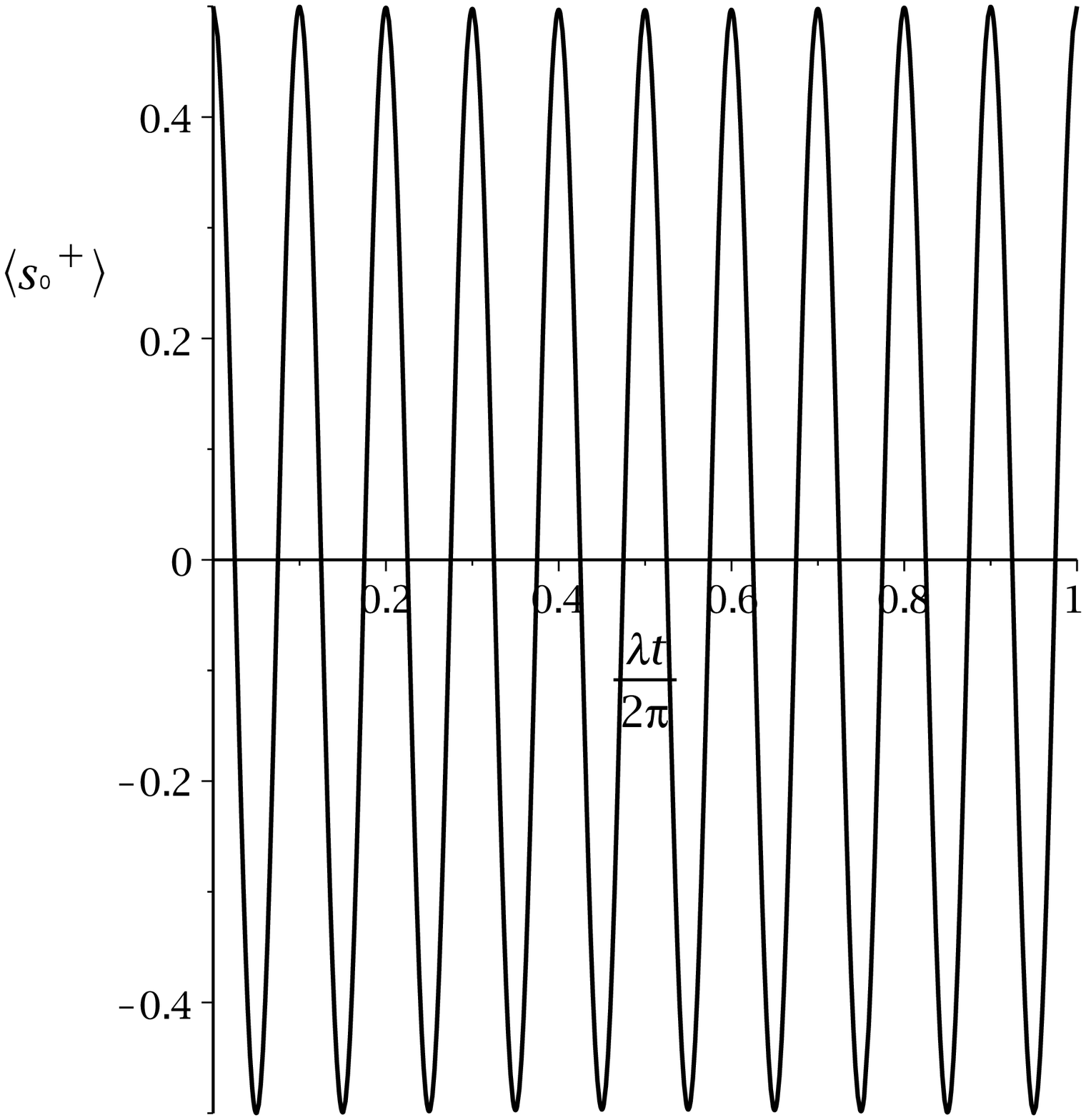}}
\caption{The connection between the Lee-Yang zeros of ten spins $1$ Ising ferromagnet with nearest-neighbor interaction and mean value of the $s_0^x$ operator
of the probe spin-$1/2$ (\ref{averagevalue}) with partition function (\ref{averagevaluennim}). The results are presented for different temperatures: (a) $T=\infty$, (b) $T=8J$, (c) $T=2J$ and (d) $T=J/4$.}
\label{nnimtenspin}
\end{figure}

Finally, let us consider the bath which consists of $N$ spins described by the 1D ferromagnetic Ising model with nearest-neighbor interaction $J$
and which has the form of a ring. The probe spin is placed in the center of this ring and interacts with each spin of the bath (see fig. \ref{spinringcluster}).
Then, in this case the partition function takes the form
\begin{eqnarray}
Z\left(\beta,h\right)=\sum_{m_1, m_2, \ldots ,m_N=-s}^{s}e^{\beta J\sum_{i=1}^N m_im_{i+1}+\beta h\sum_{i=1}^{N} m_i},
\label{averagevaluennim}
\end{eqnarray}
where $m_i$ is the projection value of the $i$th spin on the $z$-axis.

The effective spin ring with nearest-neighbor interaction can be prepared on trapped ions \cite{SchrodCat1,EQSSTI,QSDEGHTI} or ultracold atoms
\cite{opticallattice1,opticallattice2,opticallattice3,opticallattice4}. In paper \cite{sgssrwcsna} was presented the method for simulation of the ground
state of spin ring with cavity-assisted neutral atoms. Here we consider the example where the bath consists of ten spins $1$
which interact with the probe spin-$1/2$. The exact solution of the 1D Ising model for a spin-$1$ system can be obtained through the transfer matrix method
(see, for example, \cite{baiilmmfefes}). The results for different temperatures are shown in fig. \ref{nnimtenspin}.

\section{Conclusions}

We considered the evolution of the probe spin of an arbitrary value $s_0$ under the influence of a bath composed by $N$ arbitrary spins. The spin bath is
defined by a general Ising model or isotropic Heisenberg model with ferromagnetic interaction and at the initial moment of time is in thermodynamic equilibrium. The interaction
between the probe spin and each spin of the bath is defined by the Ising model. We found the relation between the observed values of the probe
spin and the partition function of the bath with complex magnetic field (\ref{averagevalue}). We showed that vanishing of these
values corresponds to the Lee-Yang zeros of the bath partition function. Thus, measuring the mean values of the probe spin as a function of time,
such as magnetization or susceptibility, we can detect the Lee-Yang zeros of the bath. This fact allowed us to obtain the connection between the moments of time
when the mean values of the probe spin vanishes and the Lee-Yang zeros of the bath.

We examined the connection between the measured values of the probe spin and the Lee-Yang zeros of the bath for some models. Namely, we considered the
triangle spin cluster, where the bath consists of two spins $s$, and the third spin-$s_0$ is a probe. The interaction between the all spins described
by Ising Hamiltonian (\ref{triangleham}). So, we obtained the correspondence between the magnetization of probe spin
and Lee-Yang zeros of bath in the case of $s=s_0=5/2$ (fig. \ref{LYZTSC}). We proposed to apply these results to the physical system of manganese
ferrite (MnFe$_2$O$_4$), where the bath and probe spin consists of two ions of Fe$^{3+}$ and ion of Mn$^{2+}$, respectively. Also the obtained results
can be apllied to the following ferrites:  ZnFe$_2$O$_4$, CoFe$_2$O$_4$, NiFe$_2$O$_4$, where the roles of the probe spins play ions of Zn$^{2+}$,
Co$^{2+}$, Ni$2+$, respectively. Finally, we studied the connection between the Lee-Yang zeros of long-range Ising bath and 1D Ising bath with nearest-neighbor interaction,
and mean values of probe spin, respectively.
In the case of long-range Ising model we considered the implementation of obtained results on the nuclear spins of tetrahedron molecules such as CH$_4$, SiH$_4$,
GeH$_4$ and SnH$_4$ (fig. \ref{lritispin1}). In the case of 1D Ising bath the connection between the Lee-Yang zeros and the magnetization of probe spin was obtained
for ten-spin ring bath prepared on trapped ions or ultracold atoms (fig. \ref{nnimtenspin}).

\section{Acknowledgments}
The authors thank Dr. Taras Verkholyak and Prof. Andrij Rovenchak for useful comments. This work was supported in part by Project FF-30F (No. 0116U001539) from the Ministry of Education and Science of Ukraine and
by the State Fund for Fundamental Research under the project F76.


\begin{thebibliography}{99}
\bibitem{LeeYangZeros1} C. N. Yang, T. D. Lee, Phys. Rev. {\bf 87}, 404 (1952).
\bibitem{LeeYangZeros} T. D. Lee, C. N. Yang, Phys. Rev. {\bf 87}, 410 (1952).
\bibitem{FisherZeros} M. E. Fisher, {\it in Lectures in Theoretical Physics, editet by W. E. Brittin} (University of Colorado Press, Boulder, CO, 1965),
Vol. 7c, p.1.
\bibitem{PYSM} F. Y. Wu, Int. J. Mod. Phys. B {\bf 22}, 1899 (2008).
\bibitem{genforarbspin1} T. Asano, Prog. Theor. Phys. {\bf 40}, 1328 (1968).
\bibitem{genforarbspin2} M. Suzuki, J. Math. Phys. (N.Y.) {\bf 9}, 2064 (1968).
\bibitem{genforarbspin3} R. B. Griffiths, J. Math. Phys. (N.Y.) {\bf 10}, 1559 (1969).
\bibitem{genforarbint1} M. Suzuki, Prog. Theor. Phys. {\bf 40}, 1246 (1968).
\bibitem{genforarbint2} M. Suzuki, M. E. Fisher, J. Math. Phys. (N.Y.) {\bf 12}, 235 (1971).
\bibitem{genforarbint3} D. A. Kurtze, M. E. Fisher, J. Stat. Phys. {\bf 19}, 205 (1978).
\bibitem{genforarbint4} I. Bena, M. Droz, A. Lipowskil, Int. J. Mod. Phys. B {\bf 19}, 4269 (2005).
\bibitem{LeeYangZerosahi} T. Asano, Phys. Rev. Lett. {\bf 24}, 1409 (1970).
\bibitem{zerospartfuncspin1} Xinhua Peng, Hui Zhou, Bo-Bo Wei, Jiangyu Cui, Jiangfeng Du, Ren-Bao Liu, Phys. Rev. Lett. {\bf 114}, 010601 (2015).
\bibitem{zerospartfuncspin2} Ch. Binek, Phys. Rev. Lett. {\bf 81}, 5644 (1998).
\bibitem{zerospartfuncspin3} Bo-Bo Wei, Ren-Bao Liu, Phys. Rev. Lett. {\bf 109}, 185701 (2012).
\bibitem{zerospartfuncspin4} Bo-Bo Wei, Shao-Wen Chen, Hoi-Chun Po, Ren-Bao Liu, Sci. Rep. {\bf 4}, 5202 (2014).
\bibitem{zerospartfuncspin5} M. Krasnytska, B. Berche, Yu. Holovach, R. Kenna, EPL {\bf 111}, 60009 (2015).
\bibitem{zerospartfuncspin6} M. Krasnytska, B. Berche, Yu. Holovach, R. Kenna, J. Phys. A {\bf 49}, 135001 (2016).
\bibitem{zerospartfuncspin7} Kh. P. Gnatenko, A. Kargol, V. M. Tkachuk, Physica A {\bf 509}, 1095 (2018).
\bibitem{zerospartfunbose1} O. Mulken, P. Borrmann, J. Harting, H. Stamerjohanns, Phys. Rev. A {\bf 64}, 013611 (2001).
\bibitem{zerospartfunbose2} W. van Dijk, C. Lobo, A. MacDonald, R. K. Bhaduri, Can.  J. Phys. {\bf 93}, 830 (2015).
\bibitem{zerospartfunbose3} P. Borrmann, O. Mulken, J. Harting,  Phys. Rev. Lett. {\bf 84}, 3511 (2000).
\bibitem{zerospartfunbose4} Kh. P. Gnatenko, A. Kargol, V. M. Tkachuk, Phys. Rev. E {\bf 96}, 032116 (2017).
\bibitem{zerospartfunbose5} Kh. P. Gnatenko, A. Kargol, V. M. Tkachuk, EPL {\bf 120}, 30004 (2017).
\bibitem{zerospartfunfermi1} R. K. Bhaduri, A. MacDonald, W. van Dijk, EPL {\bf 96}, 56003 (2011).
\bibitem{zerospartfunfermi2} A. A. Zvyagin, Phys. Rev. B {\bf 95}, 075122 (2017).
\bibitem{ylesdfehfmd} Ch. Binek, W. Kleemann, H. A. Katori, J. Phys. Cond. Matt. {\bf 13}, L811 (2001).
\bibitem{eddlyz} K. Brandner, V. F. Maisi, J. P Pekola, J. P. Garrahan, Ch. Flindt, Phys. Rev. Lett. {\bf 118}, 180601 (2017).
\bibitem{mnferrite1} E. W. Gorter, Philips Res. Rep. {\bf 9}, 295 (1954).
\bibitem{mnferrite2} J. M. Hasting, L. M. Corliss Phys. Rev. {\bf 104}, 328 (1956).
\bibitem{mnferrite3} J. Smit, H. P. J. Wijn, {\it Ferrites} (Wiley, New York, 1959), p. 136.
\bibitem{mnferriteexi1} D. J. Singh, M. Gupta, R. Gupta, Phys. Rev. B {\bf 65}, 064432 (2002).
\bibitem{mnferriteexi2} Xu Zuo, C. Vittoria, Phys. Rev. B {\bf 66}, 184420 (2002).
\bibitem{mnferriteexi3} Xu Zuo, C. Vittoria, J. Appl. Phys. {\bf 93}, 8017 (2003).
\bibitem{fivespincluster} A. Hinchliffe, D. B. Cook, Theoret. Chim. Acta (Berl.) {\bf 17}, 91 (1970).
\bibitem{SchrodCat1} K. Molmer, A. Sorensen, Phys. Rev. Lett. {\bf 82}, 1835 (1999).
\bibitem{EQSSTI} D. Porras, J. I. Cirac, Phys. Rev. Lett. {\bf 92}, 207901 (2004).
\bibitem{QSDEGHTI} J. G. Bohnet, B. C. Sawyer, J. W. Britton, M. L. Wall, A. M. Rey, M. Foss-Feig, J. J. Bollinger, Science {\bf 352}, 1297 (2016).
\bibitem{opticallattice1} L.-M. Duan, E. Demler, M. D. Lukin, Phys. Rev. Lett. {\bf 91}, 090402 (2003).
\bibitem{opticallattice2} A. B. Kuklov, B. V. Svistunov, Phys. Rev. Lett. {\bf 90}, 100401 (2003).
\bibitem{opticallattice3} I. Bloch, {\it Many-Body Physics with Ultracold Gases, edited by C. Salomon, G. Shlyapnikov, L. F. Cugliandolo}
(Oxford University Press, Oxford, UK, 2013), pp. 71-108.
\bibitem{opticallattice4} Ch. Gross, I. Bloch, Science {\bf 357}, 995 (2017).
\bibitem{sgssrwcsna} Peng Xue, Xiang Zhang, Zhihao Bian, Sci. Rep. {\bf 5}, 7623 (2015).
\bibitem{baiilmmfefes} J. Stre\v{c}ka, M. Ja\v{c}\v{s}cur, Acta Physica Slovaca {\bf 65}, 235 (2015).
\end{thebibliography}
\end{document}